%% file: main.tex
\documentclass[conference]{IEEEtran}
\IEEEoverridecommandlockouts
\usepackage{cite}
\usepackage{amsmath,amssymb,amsfonts}
\usepackage{graphicx}
\usepackage{textcomp}
\usepackage{xcolor}
\usepackage{enumerate}
\usepackage{enumitem}
\usepackage{dsfont}
\usepackage[linesnumbered, ruled, vlined,resetcount]{algorithm2e} 
\usepackage{utfsym}
\usepackage{bm}
\usepackage{graphics}
\usepackage{balance}
\ifCLASSOPTIONcompsoc
    \usepackage[caption=false, font=normalsize, labelfont=sf, textfont=sf]{subfig}
\else
\usepackage[caption=false, font=footnotesize]{subfig}
\fi
\usepackage{amsthm}
\usepackage{tcolorbox}
\usepackage{bm}
\usepackage{verbatim} 
\usepackage{diagbox}
\usepackage{multirow}
\usepackage{booktabs}
\usepackage{algpseudocode}
\usepackage{listings}
\usepackage{{makecell}}
\usepackage[ruled, vlined, linesnumbered]{algorithm2e}
\usepackage{wrapfig}
\usepackage[hidelinks]{hyperref} 
\def\BibTeX{{\rm B\kern-.05em{\sc i\kern-.025em b}\kern-.08em
    T\kern-.1667em\lower.7ex\hbox{E}\kern-.125emX}}

\input{commands}

\begin{document}

\title{
GeoLayer: Towards Low-Latency and Cost-Efficient Geo-Distributed Graph Stores with Layered Graph
}


\author{
Feng Yao, 
Xiaokang Yang, 
Shufeng Gong, 
Song Yu,
Yanfeng Zhang, 
Ge Yu \\
\IEEEauthorblockA{
\textit{Northeastern University}, Shenyang, China \\
\{yaofeng, yangxk, yusong\}@stumail.neu.edu.cn,  \{gongsf, zhangyf, yuge\}@mail.neu.edu.cn}%
}

\maketitle

\begin{abstract}
The inherent connectivity and dependency of graph-structured data, combined with its unique topology-driven access patterns, pose fundamental challenges to conventional data replication and request routing strategies in geo-distributed cloud storage systems.
In this paper, we propose \system,  a geo-distributed graph storage framework that jointly optimizes graph replica placement and pattern request routing. 
We first construct a latency-aware layered graph architecture that decomposes the graph topology into multiple layers, aiming to reduce the decision space and computational complexity of the optimization problem, while mitigating the impact of network heterogeneity in geo-distributed environments.
Building on the layered graph, we introduce an overlap-centric replica placement scheme to accommodate the diversity of graph pattern accesses, along with a directed heat diffusion model that captures heat conduction and superposition effects to guide data allocation.
For request routing, we develop a stepwise layered routing strategy that performs progressive expansion over the layered graph to efficiently retrieve the required data.
Experimental results show that, compared to state-of-the-art replica placement and routing schemes, \system achieves a 1.34$\times$–3.67$\times$ improvement in response times for online graph pattern requests and a 1.28$\times$–3.56$\times$ speedup in offline graph analysis performance.
\end{abstract}


\setcounter{page}{1}
\pagestyle{plain}

\input{1-Introduction}

\input{2-Background}

\input{3-Model}

\input{4-Layer}

\input{5-Replica}

\input{6-Routing}

\input{7-Experiment}

\input{8-RelatedWork}
\input{sample.bbl}



\clearpage
\balance

\bibliographystyle{IEEEtran}
\bibliography{sample}

\clearpage
\balance
\input{Appendix}

\end{document}

%% file: commands.tex
\usepackage{newfloat}
\usepackage{multirow}
\DeclareFloatingEnvironment[fileext=frm,placement={!ht},name=Frame]{myfloat}
\DeclareSubrefFormat{parens}{#1(#2)}
\makeatletter
\newcommand{\removelatexerror}{\let\@latex@error\@gobble}
\makeatother
\newcommand{\kw}[1]{{\ensuremath {\mathsf{#1}}}\xspace}

\newcommand{\eat}[1]{}
\newcommand{\eata}[1]{\eat{v0301:reduce pages}}

\newcommand{\stitle}[1]{\vspace{1.6ex}\noindent{\bf #1}}
\newcommand{\etitle}[1]{\vspace{1ex}\noindent{\underline{\em #1}}}

\newcommand{\beqn}{\begin{eqnarray*}}
\newcommand{\eeqn}{\end{eqnarray*}}

\newcommand{\ie}{\emph{i.e.,}\xspace}
\newcommand{\eg}{\emph{e.g.,}\xspace}

\newcommand{\resp}{\emph{resp.}\xspace}

\newcommand{\revise}[1]{\textcolor{black}{#1}}

\newcommand{\eop}{\hspace*{\fill}\mbox{$\Box$}}     
\newcounter{example}
\renewcommand{\theexample}{\arabic{example}}
\newenvironment{example}{
        \vspace{1.5ex}
        \refstepcounter{example}
        {\noindent\bf Example \theexample:}}{
        \eop\vspace{1.5ex}}

\newenvironment{proofS}{
        \vspace{1ex}
       {\noindent\bf Proof Sketch.\ }}{\eop\vspace{1ex}}

\newcommand{\bi}{\begin{itemize}}
\newcommand{\ei}{\end{itemize}}
        {\end{itemize}} 

\newcommand{\PageRank}{\kw{PageRank}}
\newcommand{\SSSP}{\kw{SSSP}}

\newcommand{\relu}{\kw{ReLU}}

\newcommand{\system}{\kw{GeoLayer}}

\newcommand{\Oi}{\mathcal{V}}
\newcommand{\E}{\mathcal{E}}

\newcommand{\Ht}{\mathcal{H}}

\newcommand{\Pb}{\mathcal{P}}

\newcommand{\equref}[1]{{Equation~(\ref{#1})}}

\newcommand{\figref}[1]{{Figure~\ref{#1}}}

\newcommand{\lemref}[1]{{Lemma~\ref{#1}}}
\newcommand{\algoref}[1]{{Algorithm~\ref{#1}}}

\newcommand{\yaof}[1]{{\color{blue}{#1}}}

\theoremstyle{definition}
\newtheorem{definition}{Definition}

\theoremstyle{theorem}
\newtheorem{theorem}{Theorem}

\theoremstyle{lemma}
\newtheorem{lemma}{Lemma}

\usepackage{tikz}
\newcommand{\blackcircleone}{%
\tikz[baseline=(char.base)]{
    \node[shape=circle,fill=black,draw,text=white,inner sep=0.5pt] (char) {1};
}}

\newcommand{\blackcircletwo}{%
\tikz[baseline=(char.base)]{
    \node[shape=circle,fill=black,draw,text=white,inner sep=0.5pt] (char) {2};
}}

%% file: 1-Introduction.tex
\section{Introduction}
\label{sec:intro}

\nocite{fullpaper}

Graph-structured data is widely used across various domains such as social networks~\cite{ICNP12cost}, 
intelligent transportation systems~\cite{TPDS16traffic},  financial transaction networks~\cite{financial-transaction}, and logistics systems~\cite{tnde20service}.
In many practical deployments, such data is often generated at geographically dispersed endpoints.
To deliver low-latency access to nearby clients, large organizations strategically deploy their applications across multiple \textit{geo-distributed} data centers (DCs).
\revise{For example, Meta operates more than 20 DCs globally to store and manage its large-scale social networking services across countries~\cite{Facebookdau}.}

An example is the infrastructure of global payment platforms~\cite{Alipay2025,Paypal2025}, which supports a range of regional and global financial business scenarios: 
(1) Transaction graphs and account relationship networks~\cite{FinancialGraph} continuously generated within specific regions are maintained by local DCs to ensure low-latency processing of regional operations such as fund transfers and credit-risk assessment~\cite{mitra24risk, AAAI25Credit}.
(2) These platforms also support cross-border transactions and provide access to globally distributed financial services~\cite{globalfinancial,globalfinancialmarket}, ensuring secure operations by tracking and monitoring transaction paths across regions.
(3) At the global scale, these platforms conduct graph analysis to deliver personalized financial services~\cite{financial2022}.

\begin{figure}[t!]
\vspace{-0.15in}
\centerline{\includegraphics[
    scale=0.55]{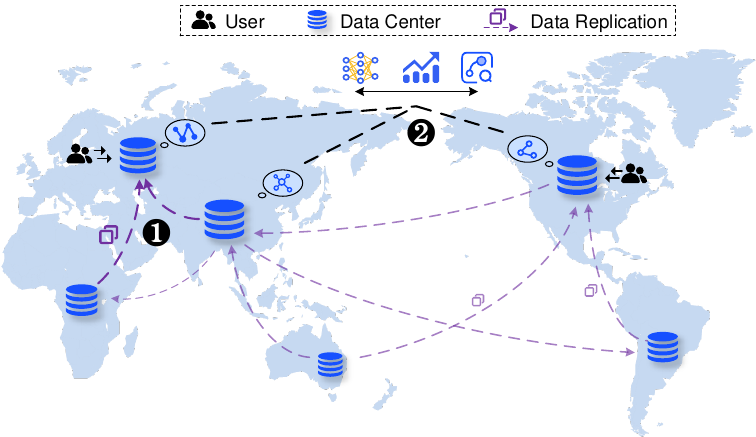}}
  \centering
  \vspace{-0.15in}
  \caption{Geo-distributed graph data storage services. }
  \label{fig:intro_example}
  \vspace{-0.25in}
\end{figure}

Geo-distributed storage services serve as the foundational infrastructure for such scenarios, offering two key capabilities: low-latency data access enabled by geo-optimized replica placement, and coordinated routing of data resources across multiple DCs.
Specifically, in cross-border transactions, the platform must perform real-time access to remote transaction graph patterns from the initiator to detect potential fraudulent activities~\cite{sigmod23geaflow}.
For example, Ant Group scans approximately 400 million potentially risky transaction patterns per day, with each risk control decision constrained to within 500 milliseconds~\cite{ant}.
However, geo-distributed DCs are connected via Wide Area Networks (WANs), where limited bandwidth often leads to significant latency for remote access.
To mitigate this, the storage service reduces access latency by either replicating pattern-related data locally or routing data requests through the nearest DCs, as shown in \figref{fig:intro_example} \blackcircleone.
In addition, \figref{fig:intro_example} \blackcircletwo \ illustrates a global-scale analytical workflow, in which the required graph data is retrieved and routed to three geo-distributed DCs.
Enhancing service availability and reliability fundamentally relies on effective replica placement and request routing,  yet these problems are non-trivial.



Service providers require that applications satisfy certain latency constraints while minimizing monetary costs, including both data transfer and storage expenses.
However, the limited and costly bandwidth of WANs introduces a critical trade-off: replicating data across multiple DCs can significantly reduce remote access latency, but it also introduces additional inter-DC synchronization traffic to maintain consistency among replicas and increased storage costs.
Furthermore, when a request spans multiple remote data items or involves global analysis, the presence of multiple replicas across DCs introduces a routing decision problem, as any of the available replicas may serve the request~\cite{INFOCOM15ADP}.
An effective routing strategy should minimize remote data access latency by selecting the optimal replica set, while simultaneously reducing inter-DC transfer costs by consolidating data access across fewer locations.




Most existing solutions for replica placement and request routing either focus solely on optimizing the distance between users and data items while overlooking the associations among data units within a request~\cite{arxiv25skystore,ICDCS21geocol,NSDI15costlo,JSAC19writeaware,kbs21adaptive,sigmod24skypie,TCC17cost}, or rely on job-specific optimizations based on predefined access patterns~\cite{infocom14multi,INFOCOM15ADP,infocom16sketch,tpds20scalable, TSC20ADP,vldb19yugong}. 
This \textit{job-driven}~\cite{vldb19yugong,vldb24reslake} paradigm inherently assumes fixed and homogeneous data access behaviors, making it feasible to co-locate replicas associated with the same access pattern based on simple statistical analysis of historical access traces.
In contrast, associations among data units in graph-structured data are intrinsic to the graph topology. Graph workloads are typically accessed via structural traversals, which are fundamentally \textit{data-driven}, resulting in highly diverse and dynamic associations among data items.
This fundamental shift in access characteristics significantly complicates the problem space and renders existing approaches ineffective.

\vspace{-0.05in}
\stitle{Challenges.} 
We present a detailed summary of the unique challenges introduced by graph-structured data.

\textit{Graphs exhibit strong connectivity and dependency relationships}. This implies that a single request may access a large volume of associated data, often triggering substantial cross-DC traffic.
As a result, the replica placement strategy must carefully account for the topological co-location of associated data items,  substantially increasing the complexity of both replica placement and associated data routing optimization.

\textit{Graph data access exhibits substantial pattern diversity}. 
\revise{Queries often attach variables on vertex and edge attributes to different parts of the query pattern~\cite{sigmod22GPML}.
In real-world financial systems, such as Ant Group’s Alipay platform~\cite{Alipay}, each transaction records dozens of attributes that are directly relevant to queries~\cite{WWW20fraud,TIST19fraud}
(\eg counterparty information, amount, timestamp, payment method, risk score).
These attributes collectively define rich graph patterns that closely reflect real query workloads.
Furthermore, the variability of source and destination vertices specified by different queries further amplifies the diversity of access patterns and the unpredictability of access behaviors.}
In addition, data accesses frequently overlap across different patterns, \ie the same data appears in multiple patterns.
Such overlap introduces significant intra-pattern variation in access frequency, further exacerbating the limitations of traditional replica placement strategies that operate at the granularity of entire patterns.

\textit{Geo-distributed DCs typically exhibit heterogeneous network bandwidth}~\cite{TNSM24survey,SIGCOMM15low}, which leads to significant latency variation when accessing associated data items involved in a graph pattern across multiple DCs.
This network asymmetry often turns the slowest-responding DC (\ie \textit{straggler}) into the performance bottleneck~\cite{tdos20aaap}.
As a result, decision-making becomes more complex, as it must jointly account for both data locality and latency heterogeneity. 

\vspace{-0.05in}
\stitle{GeoLayer.} 
To tackle these challenges, we propose \system, a geo-distributed graph storage framework that optimizes replica placement and request routing from a data-driven perspective, leveraging the structural characteristics of the graph to guide decision-making.
\system focuses on optimizing both latency and cost in geo-distributed graph storage services.
To this end, 1) \textbf{we define system-level cost metrics and formulate a joint optimization problem} that determines both replica placement and request routing.
The problem is inherently a multi-objective, multi-constraint combinatorial challenge, characterized by a vast decision space and strict latency requirements. 
To reduce complexity, 2) \textbf{we construct a layered graph architecture} that organizes the graph topology into multiple layers based on ascending request latency. 
This design decomposes the global optimization problem into a hierarchical tree of subproblems, significantly shrinking the decision space.
Meanwhile, it mitigates the impact of network heterogeneity by confining optimization to regions within each layer that exhibit uniform latency.
Built on the layered graph, 3) \textbf{we propose an overlap-centric replica placement scheme} that performs layer-by-layer data sinking guided by replication gain. 
Inspired by the principles of heat conduction and superposition in thermodynamics, we introduce a Directed Heat Diffusion model to support data placement at the granularity of overlapping pattern regions. 
This model quantifies the affinity between placement regions and the associated data items.
For request routing, 4) \textbf{we design a stepwise layered routing strategy}. In the online mode, routing follows a bottom-up, expanding retrieval process. In contrast, the offline mode consists of a top-down global data localization phase followed by a bottom-up data assembly phase, which progressively refines the data distribution based on communication cost.


In summary, we make the following key contributions.
\begin{itemize}[leftmargin=*]
    \item We introduce cost metrics and formulate the problem of minimizing cost while satisfying latency constraints (\S \ref{sec:model}).
    
    \item We construct a latency-aware layered graph that decomposes the global optimization problem into a hierarchical tree of subproblems, effectively reducing the decision space (\S \ref{sec:layer}).

    \item We propose an overlap-centric replica placement scheme that guides progressive data sinking within the layered graph, and further introduce a directed heat diffusion model to determine overlapping pattern regions allocation (\S \ref{sec:replica}).
    
    \item We design a stepwise layered routing approach for both online and offline workloads (\S \ref{sec:routing}).
    
\end{itemize}

%% file: 2-Background.tex
\section{Background}\label{sec:background}
\vspace{-0.1in}

\stitle{Graph pattern.}  Consider a connected {\em graph} $G = (V, E)$, where $V$ is a finite set of vertices and $E \subseteq V\times V$ is a set of edges. 
A graph pattern $p$ is defined as a tuple of path patterns which, when applied to $G$, yields a set of path bindings~\cite{sigmod22GPML}. A path is an alternating sequence of vertices and edges, $ph = (v_0,e_0,v_1,e_1,\dots,e_{n-1},v_n)$, where each $v_i \in V$ and $e_i \in E$.
For ease of presentation, we overload the term \textit{pattern} to refer to the actual set of data items matched by a graph query.

\stitle{Geo-Distributed Data Center.}
We highlight network heterogeneity among geo-distributed DCs.
To illustrate this characteristic, we conduct a measurement study across five of Alibaba Cloud's geo-distributed DCs, deploying \kw{ecs.c5.xlarge} instances in North America (US East--Virginia, US West--Silicon Valley), Europe (London, UK), and the Asia-Pacific region (Singapore; Beijing, China).
Specifically, we measure the available bandwidth and round-trip time (RTT) latency between instances across DCs, as summarized in Table \ref{tab:WAN}.
The measured bandwidth ranges from 96.8 Mbps to 42.4 Mbps, while latency varies from 69.1 ms to 256.3 ms. 
This variability is primarily attributed to differences in network infrastructure, physical distance, and network topology among the DCs.

\begin{table}[t!]
\vspace{-0.15in}
\centering
\caption{Available bandwidth (BW) and RTT latency (Lat.) among Alibaba Cloud's geo-distributed DCs.}\label{tab:WAN}
\vspace{-0.1in}
\begin{tabular}{l|c|c|c|c|c}
\toprule
\diagbox{{Lat.}}{BW} & {US East}  & {US West}  & {London}   & {Singapore} & {Beijing}  \\
\midrule
US East   & {---}      & 96Mbps & 92Mbps   & 66Mbps  & 68Mbps   \\ \hline
US West   & 69ms   & {---}     & 93Mbps & 80Mbps    & 77Mbps \\ \hline
London    & 80ms    & 136ms & {---}     & 74Mbps  & 42Mbps \\ \hline
Singapore & 225ms  & 178ms & 213ms & {---}      & 96Mbps \\ \hline
Beijing   & 226ms   & 145ms & 256ms & 75ms  & {---} \\
\bottomrule
\end{tabular}
\vspace{-0.1in}
\end{table}

\begin{table}[t!]
\vspace{-0.1in}
\centering
\caption{The costs of cloud storage services offered by different cloud providers.}\label{tab:oss}
\vspace{-0.1in}
\begin{tabular}{lcccc}
\toprule
Cloud   Provider    & \begin{tabular}[c]{@{}c@{}}Storage\\ (\$/GB/Month)\end{tabular} & \begin{tabular}[c]{@{}c@{}}GET/PUT\\ (\$/M)\end{tabular} & \begin{tabular}[c]{@{}c@{}}Data Transfer\\ (\$/GB)\end{tabular} \\
\midrule
Amazon~\cite{amazonOss}  & 0.023  & 0.40 / 5.00   & 0.050  \\
Google~\cite{googleOss} & 0.020  & 0.40 / 5.00   & 0.080  \\
IBM~\cite{IBMOss}      & 0.023  & 0.42 / 5.20   & 0.050  \\
Azure~\cite{AzureOss}    & 0.021  & 0.50 / 6.50   & 0.040  \\
Alibaba~\cite{AlibabaOss}  & 0.016  & 0.10 / 1.40   & 0.043  \\ 
\bottomrule
\end{tabular}
\vspace{-0.2in}
\end{table}

\stitle{Cloud Storage Cost.} Cloud providers offer \textit{disaggregated} storage as a service in geo-distributed DCs.
Data is stored in immutable blocks called \textit{objects}, which are distributed and replicated across multiple storage nodes to ensure availability and durability~\cite{vldb23AnyBlob}.
The relevant costs primarily include storage, data retrieval and modification, and inter-region data transfer. 
Taking the US East region as an example, Table \ref{tab:oss} presents the pricing details from various providers under a peak-demand scenario.
Cloud storage services allow users to retrieve and modify data through \kw{PUT} and \kw{GET} operations, with intra-region API calls billed based on the number of requests. 
However, these services do not provide intelligent replica placement or request routing strategies.


\stitle{System Services Overview.} 
We outline the system storage services responsible for handling user-requested data items involved in queried graph patterns.

\etitle{Graph query mode.} 
Based on real-world business requirements, graph queries can be categorized into online and offline modes~\cite{CIKM16giraphasync}. 
Accordingly, the storage service exhibits significantly different response characteristics to graph pattern requests when providing data access for these two scenarios:

(1) 
\revise{Online requests typically involve both read and write operations over graph patterns. 
For read requests, query patterns commonly involve graph pattern matching (often join-heavy on many-to-many edges with group-by, aggregation, and sorting operators) and finding paths (\eg cheapest/shortest or weighted paths)~\cite{VLDB22ldbc,qi2023ldbc}.
Queries attach variables about vertex/edge attributes on different parts of the pattern~\cite{sigmod22GPML}.
For example, in a funds-transfer scenario, a read request retrieves transfers during a given period from a blocked account to an unblocked account, with variables covering the vertex type \texttt{Account}, vertex attribute \texttt{isBlocked}, edge type \texttt{Transfer}, and edge attribute \texttt{date}.
For write requests, the pattern performs insert/update/delete on involved vertices, edges, or attributes. For instance, a scheduled transfer inserts a \texttt{Transfer} edge from the sender to the receiver and updates the corresponding account balance within the specified time window.}
In online mode, graph pattern requests are primarily used in interactive applications that demand fast responses, imposing strict latency requirements.
To meet this, an optimized replica placement strategy can provide low-latency access.

(2) Offline requests primarily focus on graph analytics tasks, such as iterative graph computation~\cite{VLDB23ragraph,SOCC22pgpregel} and graph learning~\cite{zeng2022gnn,Graphlearning}. 
It is often necessary to correlate data from multiple DCs to derive global insights.
\revise{For instance, in a cross-border e-commerce scenario, the system extracts and associates user shopping histories from geo-distributed DCs and conducts a joint analysis across them to generate product recommendations.}
In offline mode, graph data requests typically have no real-time requirements but require coordination of distributed data resources across DCs to support the efficient execution of analytical algorithms, while minimizing data transmission costs.

\etitle{Data replication and request routing.}
\revise{As discussed in the online mode, requests target graph patterns, such as retrieving all transfers within a given time window from blocked to unblocked accounts. Because the source and destination accounts (and their related transactions) may reside in different DCs, low-latency access is achieved by on-demand replicating the pattern-related items to the DC closest to the user and routing the request locally whenever possible.
During workload execution, access logs are continuously collected to capture per-item request frequencies, which are then used to guide adaptive replication decisions~\cite{tpds20scalable}. 
In the offline mode, existing replicas remain reusable. For joint analytics, request routing selects an optimal layout over the available replicas, favoring in-place computation and locality-preserving join, to reduce inter-DC data transfer.
}

%% file: 3-Model.tex
\section{FORMULATION AND MODELING}\label{sec:model}
\subsection{System Model}\label{subsec:system_model}
We first introduce the definition of request latency, then present the \system's cost metrics.

\stitle{Request Latency.} 
We define request latency as a measure of pattern request execution efficiency, which comprises the RTT latency incurred by TCP connection establishment (denoted by $l^{RTT}$) and the latency of data item transmission~\cite{ICDCS21geocol}. 
The request latency $l_{yd}^p$ for a request $p$ originating from DC $y$ that is responded to by the DC $d$ is computed as follows:
\begin{equation}
\begin{aligned}
   \small l_{yd}^p = l^{RTT}_{dy}  + {S^p_d}/{BW_{dy}},
\end{aligned}
\end{equation}
where $S_d^p$ denotes the size of the data items involved in $p$ that reside at DC $d$, and $BW_{dy}$ represents the available bandwidth between DC $d$ and DC $y$.

Since data item requests within request $p$ may span multiple DCs,
we assume that $p$ involves $n$ DCs.
The overall request latency $l^p_y$ is defined as the maximum access latency across all participating DCs, \ie $l^p_y = \max_{1\leq d \leq n} \{l^p_{yd}\}$.
The DCs involved in serving request $p$ respond in parallel, after which the partial results are assembled into the final response~\cite{VLDB14assemble}.
However, due to network heterogeneity, some workers may experience substantially higher latency than others. The worker with the highest latency, commonly referred to as the \textit{straggler}, becomes the main bottleneck and ultimately determines the overall latency~\cite{tdos20aaap}.


\stitle{Cost Metrics.} We formulate the major decision factors.

\etitle{Storage cost.} 
The storage cost quantifies the expense of storing data items and their replicas across all DCs. It is defined as:
\begin{equation}\label{eq:store_cost}
\begin{aligned}\small
     C^{(S)} = \sum_{x \in I} s_x \sum_{d \in D} \delta_{xd} \cdot c_d^{store},
\end{aligned}
\end{equation}
where $I$ is the set of data items, and $s_x$ is the size of item $x$. $\delta_{xd}$ is a binary variable that equals 1 if $x$ is stored in DC $d$, and 0 otherwise; $c^{store}_d$ denotes the unit storage cost in DC $d$.

\etitle{Request cost.} The request cost accounts for both read (GET) and write (PUT) operations. We denote $R_{xy}$ (\resp $W_{xy}$) as the frequency of read (\resp write) requests for data item $x$ originating from DC $y$ over a given time window. 

For read requests, beyond the GET cost ($c_d^{read}$), accessing non-local data items incurs additional network costs due to cross-DC transfers.
The cost of read requests is defined as:
\begin{equation}
\begin{aligned}
\small 
   C^{(R)} =  \sum_{x \in I}\sum_{y \in D} R_{xy}  \sum_{d \in D}\sigma_{xyd}  \Big( c_d^{read} + \mathbb{I}(y \neq d)\cdot s_x \cdot c_{dy}^{net}\Big), 
\end{aligned}
\end{equation}
where $\sigma_{xyd}$ indicates whether DC $d$ provides access to data $x$ requested by DC $y$, and  $\mathbb{I}(y \neq d)$ is an indicator function that equals 1 for cross-DC requests ($y \neq d$) and 0 otherwise.
$c_{dy}^{net}$ denotes the unit data transmission cost from DC $d$ to DC $y$.

The cost of write requests encompasses the expenses incurred from PUT cost and the network costs associated with synchronizing updates across all replicas, which can be quantified as follows:
\begin{equation}
\begin{aligned}
\small C^{(W)} = \sum_{x \in I}\sum_{y \in D} W_{xy}  \left[c^{write}_{y}+\sum_{\substack{d \in D \\ d \neq y}}\delta_{xd} \Big(c^{write}_{d} + s_x \cdot c_{yd}^{net}\Big)\right],
\end{aligned}
\end{equation}
where $c^{write}_{y}$ is the PUT cost of the originating DC and $c^{write}_{d}$ is the synchronized PUT cost.

\etitle{Association penalty cost.} 
Data items associated with a graph pattern may reside across multiple DCs.
In distributed systems, serving such read pattern requests can incur additional system overhead,
\ie coordination costs incurred by distributed routing, which tends to increase with the number of participating DCs~\cite{INFOCOM15ADP}.
In addition, as previously discussed, the straggler in heterogeneous networks is often the bottleneck for request latency.
Considering both factors, \system incorporates an association penalty cost component to promote the co-location of associated data items at a minimal number of storage sites, while avoiding routing across highly heterogeneous network paths.
The association penalty cost is given by:
\begin{equation}\label{eq:cost_function}
\begin{aligned}
   &\small C^{(A)} = \sum_{p \in P}\sum_{y \in D} R_{py} \left[\lambda_1\bigg(\sum_{d \in D}\rho_{pyd} - 1\bigg) + \lambda_2\Delta l^p_{y}  \right], \\
   &\small \text{where }\quad \Delta l^p_{y} = \frac{\max\limits_{\{y,d \in D \mid \rho_{pyd}= 1\}} l^p_{yd} -\min\limits_{\{y,d \in D \mid \rho_{pyd}= 1\}} l^p_{yd}}{\min\limits_{\{y,d \in D \mid \rho_{pyd}= 1\}} l^p_{yd}}.
\end{aligned}
\end{equation}

Here, $\lambda_1$ and $\lambda_2$ are penalty coefficients, set to 0.5 in our experiments.
$\rho_{pyd}$ indicates whether DC $d$ is used to serve at least one data item in the pattern $p$ originated by DC $y$.
Ideally, if all data items in a pattern are located within one DC, \ie $\sum_{d \in D}\rho_{pyd} = 1$, no penalty is incurred. 
Otherwise, the penalty grows with the number of participating DCs involved in serving the pattern.
Additionally, the maximum latency between DCs involved in pattern $p$ (\ie straggler) should not be excessively disparate.
\equref{eq:cost_function} reflects \system's preference for data locality access and latency uniformity.

\subsection{Optimization Problem} 
\system's objective is to minimize the overall cost by determining  1) the placement of data items involved in patterns, \ie $\delta_{xd}$, and 2) the routing of pattern read requests, \ie $\sigma_{xyd}$ and $\rho_{pyd}$, while satisfying all constraints:
\begin{equation}\label{eq:objective_func}
\begin{aligned}\small 
&\min_{\delta_{xd},\ \sigma_{xyd},\ \rho_{pyd}}\Big(C^{(S)}+ C^{(R)} +C^{(W)} + C^{(A)}\Big), \\
\end{aligned}
\end{equation}
\[\begin{aligned}
\text{s.t.} \quad
&\small \text{(a)} \quad  \sigma_{xyd} \leq \delta_{xd}, \ \sum_{d \in D} \sigma_{xyd} = 1, \\
&\small \text{(b)} \quad \rho_{pyd} \leq \delta_{x^*d}, \ \forall x^* \in I_p, \\
&\small \text{(c)} \quad \frac{1}{|I|} \sum_{x \in I}\sum_{y \in D} R_{xy} \sum_{d \in D} \sigma_{xyd} \cdot l^x_{yd} \leq \Gamma_{\max},\\
&\small \text{(d)} \quad \max_{\{y,d \in D \mid \rho_{pyd}= 1\}}  l^p_{yd} \leq \eta_p \cdot \Gamma_{\max}, \\
&\small \text{(e)} \quad \delta_{xd},\ \sigma_{xyd}, \ \rho_{pyd} \in \{0,1\}, \ \forall x \in I, \forall y,d \in D, \forall p \in P. \\
\end{aligned}
\]

Here, constraint (a) ensures that each data item is routed exclusively to a DC that holds its replica, and that each request for a data item is served by a single DC.
Constraint (b) ensures that read requests for data items in pattern $p$ are routed only to the DCs storing the corresponding replicas, where $I_p \subseteq I$ denotes the set of data items associated with pattern $p$.
Moreover, \system emphasizes adherence to latency Service Level Objectives (SLOs).
Let $\Gamma_{\max}$ denote the upper bound on acceptable latency. Constraint (c) enforces that the average response latency of read requests does not exceed $\Gamma_{\max}$.
Constraint (d) further stipulates that the maximum latency for each request pattern $p$ must remain below $\eta_p \cdot \Gamma_{\max}$, where $\eta_p \in (0, 1]$ is a proportionality coefficient that reflects the varying latency requirements of different query patterns. For instance, fraud detection in financial transactions often demands rapid responses (\eg $\leq$ 500ms), whereas logistics delivery path queries can tolerate more relaxed thresholds~\cite{Eur02modeling}.


The optimization problem can be formulated as a binary integer programming (BIP) model, which has been proven to be NP-hard~\cite{82nphard}.
The decision variables in this model are tightly interdependent. 
Replica placement directly constrains the feasible space for routing decisions, requiring the co-location of associated data items involved in the graph pattern and necessitating dynamic trade-offs between request latency and cross-DC synchronization overhead.
Simultaneously, routing decisions must jointly consider data locality, based on current replica distribution, and end-to-end request latency. 


%% file: 4-Layer.tex
\section{Latency-aware layered graph}\label{sec:layer}

\stitle{Overview.}
The optimization problem is a multi-objective, multi-constraint combinatorial task for which direct global optimization is computationally intractable.
We address this by constructing a latency-aware layered graph that decomposes the global problem into a hierarchical tree of subproblems. 
Patterns involving a large number of \textit{cross-partition edges} inherently trigger inter-DC data accesses~\cite{ICDE22adaptive}, with request latencies varying significantly due to network heterogeneity.
We organize these cross-partition edges into multiple layers based on ascending request latency. By construction, inter-layer latencies progress monotonically.
Within each layer, a set of edges with similar response latencies groups the connected lower-layer graph regions into latency-homogeneous clusters.
The connections among these layer-wise edge sets and clusters form a hierarchical tree. \revise{Restricting placement and routing decisions to the branches of this tree substantially reduces the decision space, mitigates variable coupling, and provides a unified, stable structure for subsequent overlap-centric replica placement (\S\ref{sec:replica}) and stepwise layered routing (\S\ref{sec:routing}).}

\stitle{Layered Graph Construction.} 
Given a \textit{connected graph} $G = (V, E)$ and its \textit{geo-distributed partitioning}, the layered graph is constructed by combining three components: (1) the base layer, denoted as $Layer_0$, which consists of per-DC subgraphs $\{G_d = (V_d,E_d)\}_{d \in D}$; (2) upper-layer bridge graphs ($Layer_i, 1\leq i\leq h$), denoted as $Layer_i : G^B_i = (V^B_i, E^B_i)$ which capture inter-partition connectivity at different latency levels; and (3) the inter-layer links that connect bridge graphs across adjacent layers.


\etitle{Underlying graph ($Layer_0$).}
\revise{In geo-distributed environments, each DC independently stores and maintains a local subgraph. This design isolates locally available data without requiring WAN access and constitutes the base layer ($Layer_0$) of the layered graph, representing the local substructures of $G$ maintained by individual DCs.}
Formally, the base layer $Layer_0$ consists of $|D|$ disjoint subgraphs $\{G_d = (V_d,E_d)\}_{d \in D}$, obtained by geo-distributed partitioning of $G$. Each $G_d$ is stored at DC $d$ and satisfy:
(i) $\bigcup_{d \in D} V_d=V$; (ii) $V_d\cap V_{d'}=\emptyset$ for $\forall d\neq d'$; (iii) $E_d = E \cap (V_d \times V_d)$. 


\etitle{Upper-layer bridge graph ($Layer_i, 1\leq i\leq h$).} 
\revise{The \emph{bridge graph} is constructed from cross-partition edges that connect boundary vertices across different DC subgraphs $G_d$, enabling data access across DCs. Considering the latency heterogeneity of WAN links between DCs, these cross-partition edges are assigned to different latency layers. This organization forms locally uniform-latency regions within each layer.}


\begin{definition}[Bridge  Graph]
Let each subgraph \(G_d=(V_d,E_d)\) have a boundary vertex set
\(
B_d \;=\;\bigl\{\,u\in V_d \mid \exists\,v\in V\setminus V_d,\;(u,v)\in E\bigr\}.
\)
Define the global boundary vertex set
\(
V^B \;=\;\bigcup_{d\in D}B_d,
\)
and the set of cross‐partition edges
\(
E^B \;=\;\bigl\{(u,v)\in E \mid u\in B_d,\;v\in B_{d'},\;d\neq d'\bigr\}.
\)
Then, the bridge graph is defined as
\(
G^B \;=\;(V^B,\,E^B).
\)
\end{definition}

Considering the network heterogeneity among DCs, we introduce a latency assignment function $\delta: E^B \rightarrow \mathbb{R}^+$ that maps each edge in the bridge graph to inter-DC request latency. 
For any $e=(u,v) \in E^B$, we set $\delta(e) = l^{e}_{dd'}$, where $l^{e}_{dd'}$ denotes the observed latency between DCs $d$ and DC $d'$.

We then partition the bridge graph into $h$ latency-based layers to cope with network heterogeneity. 
First, define a sequence of latency thresholds
$0 = t_0 < t_1 < \dots<t_{h-1} < t_{h}= +\infty$. 
Next, introduce the latency‐partition function $f:E^B \longrightarrow \{1,2,\dots,h\}$, where $e \in E^B$, $f(e) = i \Leftrightarrow \delta(e) \in [t_{i-1},t_{i})$.
In other words, each edge whose measured latency falls within $[t_{i-1},t_{i})$  is assigned to $Layer_i$.
We then project the global bridge graph $G^B$ onto each layer:
$E^B_i = \{e \in E^B \mid f(e) = i\}$, $V^B_i = \bigcup_{e\in E^B_i} \{u,v \mid (u,v)\in e\} \subseteq V^B$.
Thus, the bridge graph at latency level $i$ is $G^B_i = (V^B_i, E^B_i)$.



To link layers, we examine how edges in the bridge graph affect connectivity in lower-layer regions and introduce the concept of a \emph{bridge subgraph} to group intra-layer edge subsets.



\begin{definition}[Bridge Subgraph]
For a given layer \(Layer_i\) with \(i > 0\), let
\(
\mathcal{G}_{i-1} = \bigl(V,\,E_{prev}\bigr)
\),
where
\(E_{prev} =
\bigcup_{d\in D} E_d
\;\cup\;
\bigcup_{j=1}^{i-1} E^B_j
\) represents the aggregated graph composed of the underlying graph ($Layer_0$) and all bridge graphs from $Layer_1$ to $Layer_{i-1}$,
and let the weakly connected components of \(\mathcal{G}_{i-1}\) be denoted as \(WCC = \{cc_1, \dots, cc_n\}\). 
A subgraph $G^*=(V^*,E^*)$ with  
\(
E^* \subseteq E^B_i
\),
\(
V^* = \bigcup_{(u,v)\in E^*}\{u,v\}
\),
is called a \emph{bridge subgraph} if adding \(G^*\) to \(\mathcal{G}_{i-1}\) merges a subset of components 
\(\{cc'_{1},\dots,cc'_{m}\}\subseteq WCC\) into a single connected component  
\(
\hat{cc}
= \Bigl(\,\bigcup_{j=1}^m cc'_{j}\Bigr)
\;\cup\;G^*
\).
The bridge subgraphs from \(Layer_{i-1}\) that are contained within \(\hat{cc}\) collectively form a \emph{cluster} linked by \(G^*\).
\end{definition}


Notably, the DCs in $Layer_0$ are grouped into clusters by the bridge subgraphs in $Layer_1$.
Moreover, within each cluster, network resources are more evenly distributed than across disjoint regions at the same layer.



\begin{figure}[t!]
\vspace{-0.2in}
\centerline{\includegraphics[
    scale=0.48]{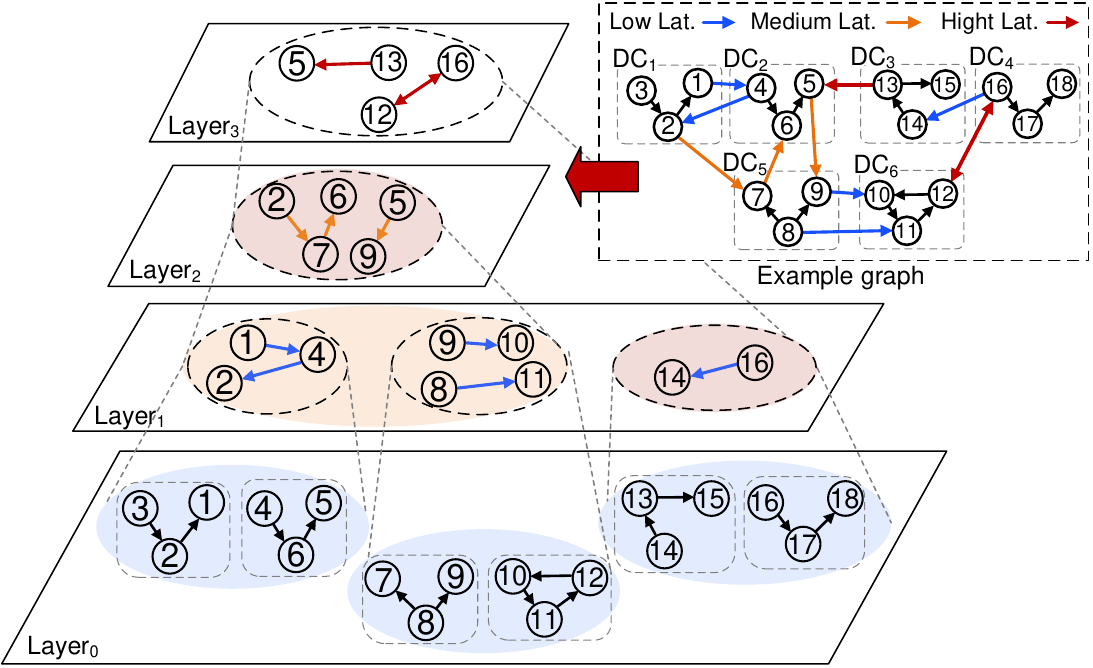}}
  \centering
  \vspace{-0.1in}
  \caption{An example of constructing a layered graph based on latency levels.}
  \label{fig:layer_graph}
  \vspace{-0.2in}
\end{figure}


\begin{example}
The right side of \figref{fig:layer_graph} depicts a geo-distributed example graph with 18 vertices spread across six DCs, \(\{DC_1,\dots,DC_6\}\). For clarity, inter-DC link latencies are abstracted into three levels, \ie low, medium, and high, and each cross-partition edge is assigned a latency level via the function \(f(e)\).
The left side of \figref{fig:layer_graph} shows the corresponding layered graph. Local subgraphs maintained by each DC form \(Layer_0\), while cross-partition edges are placed in \(Layer_1\), \(Layer_2\), and \(Layer_3\) according to their latency levels. These edges are then grouped into \emph{bridge subgraphs}, each of which merges a set of previously disjoint connected components. 
For example, in \(Layer_1\), edges \((1,4)\) and \((4,2)\) connect the subgraphs at \(DC_1\) and \(DC_2\), forming a bridge subgraph and thereby clustering \(DC_1\) and \(DC_2\). Similarly, the bridge subgraphs in \(Layer_2\) connect the clusters induced by the yellow-shaded bridge subgraphs in \(Layer_1\).
\end{example}

%% file: 5-Replica.tex
\section{Overlap-centric replica placement}\label{sec:replica}
\subsection{Design Overview} 

Building on the layered graph, our replica placement scheme localizes decision-making within clusters at each layer, operating at the pattern/overlap-region granularity.
The process begins with a sinking step that assigns each pattern to the latency-compliant layer.
To address the variation in access frequencies across overlapping regions of patterns, we first apply a gain function to determine whether a pattern should undergo full intra-cluster replication or decomposition.
For decomposable patterns, we split them along their overlapping regions and let these region‐segments competitively bid for placement across the cluster’s bridge subgraphs.
To quantify the affinity between each bridge subgraph and these segments, we introduce a directed heat diffusion model, which guides the placement of pattern segments within each cluster.
This process cascades recursively through the layers until replicas are placed in individual DCs.
\revise{By decomposing patterns at the overlap-region granularity, we weaken co-location coupling among pattern-related data and restrict placement decisions to latency-homogeneous clusters. 
Consequently, routing cost depends only on whether a cluster is covered, rendering the objective cluster-separable and enabling a place-first, then best-response routing (\S\ref{sec:routing}).}
\revise{Due to space constraints, we defer the full implementation details to~\cite{fullpaper}.}

\subsection{Replica Placement Built on Layered Graph}\vspace{-0.05in}

\stitle{Sinking of Pattern Replicas.} 
Graph pattern requests impose varying latency requirements, as reflected in constraints (d) of \equref{eq:objective_func}.
Because the layered graph is organized by increasing request latency, whenever a pattern’s latency bound cannot be met at layer $L_k$, a complete replica of that pattern must be sunk to every bridge subgraph in the corresponding clusters of $L_{k-1}$.  
In other words, the components of the pattern cannot be split across different bridge subgraph regions in layer $L_{k-1}$,  as such dispersed placement would fail to meet the latency requirements for remote access between components within the pattern.
This sinking process continues until the pattern reaches a layer that satisfies its latency requirements.
Once a layer does meet the pattern’s latency constraint, \system allows the pattern’s data items to be decomposed and distributed across the appropriate regions in all lower layers, enabling finer-grained cost optimization.
At that point, the pattern is placed at this layer in preparation for subsequent replica placement optimization.
The mechanism of sinking replicas enforces strict latency guarantees while enabling a trade-off between the cost of remote access and the benefits of local caching.


\begin{figure}[t]
\vspace{-0.35in}
    \centering
    \hspace{-0.15in}
    \subfloat[Conduction effect.]{\label{fig:path_distribution}
    \includegraphics[width=0.54\linewidth]{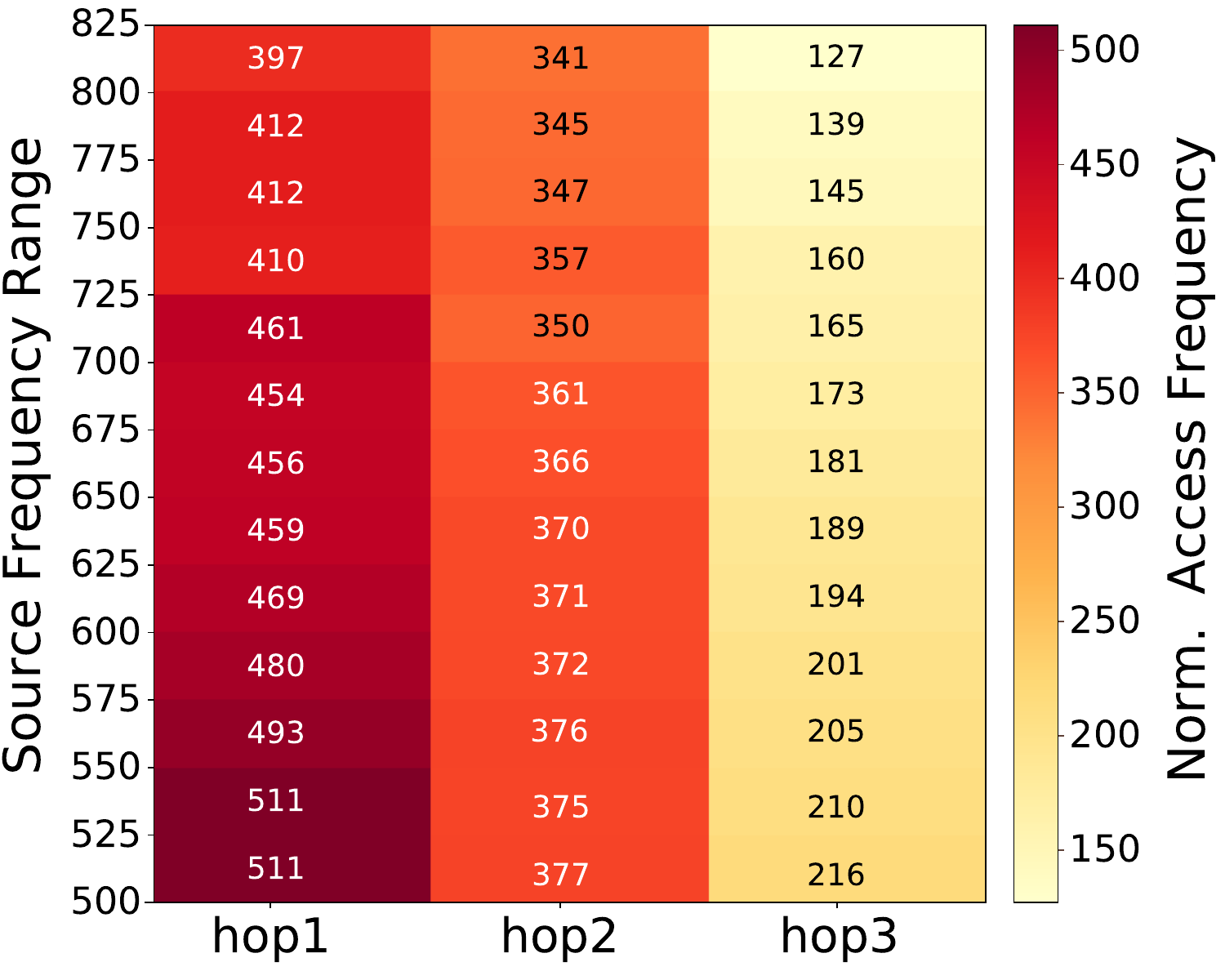}}\hspace{0.01in}
    \subfloat[Superposition effect.]{\label{fig:path_heat}
    \includegraphics[width=0.44\linewidth]{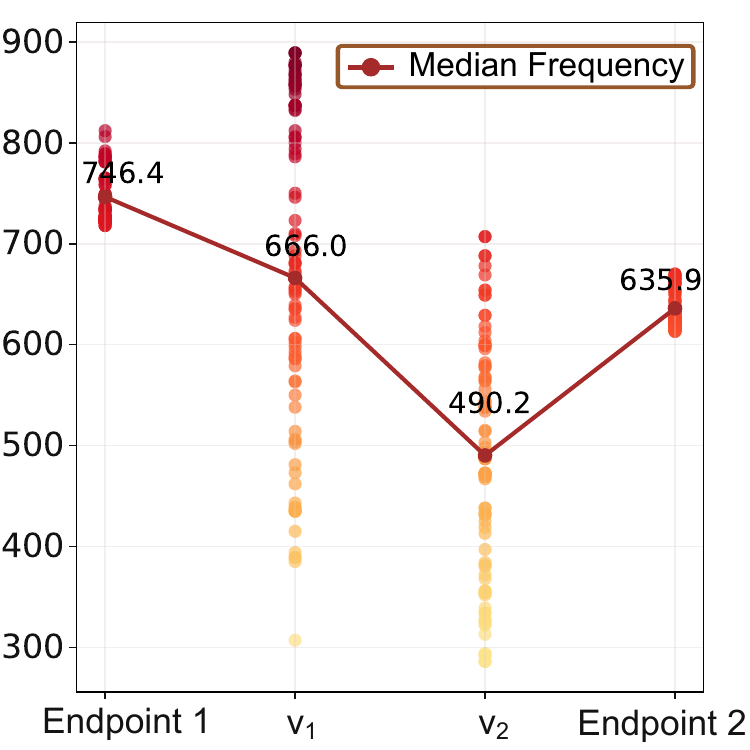}}
    \vspace{-0.05in}
    \caption{ The conduction and superposition effects of access frequency along paths in pattern access. The color intensity represents the normalized access frequency of vertices, with darker colors indicating higher frequency.}
    \label{fig:path_insight}
    \vspace{-0.2in}
\end{figure}

\stitle{Directed Heat Diffusion Model.} 
Before presenting the replica placement optimization,  we first introduce the \textit{Directed Heat Diffusion} (DHD) model.
This model captures how graph pattern access behaviors manifest over the graph topology, quantifying which regions are most likely to be involved in complete pattern requests and thus warrant co-location. 
These predictions provide a key foundation for guiding fine-grained replica placement decisions over overlapping pattern regions.



\etitle{Insight.} 
A graph pattern consists of a set of path bindings whose elements exhibit strong connectivity and dependency.
To examine how these structural characteristics affect access behavior,
we randomly issued 1,000 three-hop path queries on the LDBC dataset~\cite{LDBC} and recorded each involved vertex’s access frequency. \figref{fig:path_distribution} presents a heatmap illustrating the median access frequencies of vertices at hops 1 to 3, grouped by the access frequency range of their corresponding source vertices (y-axis).
The results reveal a clear \textit{conduction effect}: median access frequency declines steadily as path length increases.
Furthermore, vertices adjacent to high-frequency vertices themselves tend to be accessed more often. Interestingly, certain source vertices (\eg in $[500, 550]$) exhibit stronger conduction effects;  we attribute this to \textit{edge transition probabilities}, whereby a larger neighbor set dilutes the probability of traversing any individual outgoing edge.
\figref{fig:path_heat} extends this analysis to paths anchored by high-frequency endpoints and shows that intermediate vertices maintain even higher median access frequencies compared to those in \figref{fig:path_distribution}. This finding suggests that the conduction effect accumulates across hops, exhibiting a superposition behavior.


\revise{Intuitively, in real-world graphs, core regions (\eg celebrities in social networks) attract high request rates, and their influence propagates along edges to neighboring regions, elevating neighbors’ access frequencies. Overlapping influence from multiple cores superimposes at common neighbors, making their accesses even more frequent. This motivates a heat view of access: vertices act as heat sources, edges conduct heat, and the heat field dissipates unless reheated by new requests.}



\etitle{Design of the model.} Inspired by thermodynamic principles of heat conduction and superposition, we propose the Directed Heat Diffusion model, which conceptualizes vertices as thermal masses capable of absorbing and diffusing heat.

We first interpret the access frequency $R_{x\cdot}$ ($R_{x\cdot} > 0$) of each data item $x$ (including vertices and edges) as its initial heat in the system, so that $\Ht^0_x = R_{x\cdot}$.

At each diffusion step $k$, the heat transferred from vertex $u$ to vertex $v$ is:
\begin{equation}\label{eq:v_heattrans}
\begin{aligned}\small
    \Delta \mathcal{H}_{uv}^k = \alpha \cdot \frac{A_{uv}}{|\mathcal{N}^{out}_u|} \cdot \relu (\Ht^k_u - \Ht^k_v),
\end{aligned}
\end{equation}
where $A_{uv}$ in the adjacency matrix $A$ represents the initial heat value of the \textit{undirected} edge $(u,v)$.
The parameter $\alpha$ controls heat diffusivity, and the product $\alpha \cdot A_{uv}$ determines the diffusion intensity. 
The \relu operation suppresses negative values, ensuring that heat flows only from higher to lower temperature vertices.
Note that heat diffusion does not strictly follow the direction of directed edges, as pattern traversal may also involve moving towards in-neighbors~\cite{sigmod22GPML}.  The term $|\mathcal{N}^{out}_u|$ denotes the number of lower-heat neighbors.


The heat update for a vertex $v$ can be expressed as:
\begin{equation}\label{eq:v_update}
\begin{aligned}\small 
    \Ht^{k+1}_v = (1 - \gamma)\left[\Ht^k_v + \sum_{u \in \mathcal{N}^{in}_v} \Delta \Ht^k_{uv} - \sum_{w \in \mathcal{N}^{out}_v} \Delta \Ht^k_{vw}\right].
\end{aligned}
\end{equation}

Here, $\gamma$ is the global decay rate that models natural heat dissipation and captures the time-decaying nature of data access frequency due to information obsolescence~\cite{vldb19yugong,TCC17cost}. 
The sets $\mathcal{N}^{in}_v$ and $\mathcal{N}^{out}_v$ denote the neighbors from which heat flows into and out of vertex $v$, respectively.
Intuitively, heat accumulates from any higher-heat neighbor, while its heat propagates to its lower-heat neighbors. 

Additionally, certain regions $\mathcal{O}$ of the graph act as heat sources (\eg active users who frequently post content in social networks), continually supplying heat to the system. We model their contribution as source dynamics, defined as follows:
\begin{equation}\label{eq:heat_source}
\begin{aligned}\small
 Q_v^{k+1} = 
    \begin{cases}
        Q_v^0 e^{-\pi k} + \Delta Q \cdot \sum\sigma_{v\cdot \cdot}, 
            & \text{if } v \in \mathcal{O}, \\[5pt]
        0,   
            & \text{otherwise.}
    \end{cases}
\end{aligned}
\end{equation}

Here, $e^{-\pi k}$ implements exponential decay with rate $\pi = \ln(2) / T_{hl}$ determined by the half-life $T_{hl}$. We initialize each source’s heat uniformly as $Q_v^0 = 1 / |\mathcal{O}|$.
The term $\Delta Q \cdot \sum\sigma_{v\cdot \cdot}$ captures additional heat proportional to the frequency of accesses to vertex $v$.


Then the system heat evolves according to
\begin{equation}\label{eq:system_update}
\begin{aligned}\small
    \Ht^{k+1} = (1 - \gamma)\left[\Ht^k + \alpha L_{dir}^k\Ht^k\right] + \beta Q^k,
\end{aligned}
\end{equation}
where $L_{dir}$ is the directional Laplace matrix  encoding both the magnitude and direction of heat flow, defined as:
\begin{equation}
\begin{aligned}\small
    (L_{dir})_{vw} = 
    \begin{cases}
        \ -\dfrac{A_{vw}}{|\mathcal{N}^{out}_v|}, 
            & \text{if }  \Ht_v > \Ht_w, \\[10pt]
        \ \displaystyle\dfrac{A_{wv}}{|\mathcal{N}^{out}_w|}, & \text{if } \Ht_w > \Ht_v, \\[10pt]
        \ 0,   & \text{otherwise.}
    \end{cases}
\end{aligned}
\end{equation}

Finally, at equilibrium, $\Ht^{k+1} = \Ht^k =\Ht^*$, and substituting into the update \equref{eq:system_update} yields the steady-state condition
\begin{equation}\label{eq:steady_strate}
\begin{aligned}\small
    \gamma \cdot\Ht^* - \alpha (1-\gamma)\cdot L_{dir}^* \Ht^* = \beta Q^*.
\end{aligned}
\end{equation}

\revise{Regarding model parameter settings, a large $\alpha$ leads to an overly smoothed heat distribution, whereas a smaller $\alpha$ weakens propagation and slows convergence to a stable state. Therefore, a moderate value of $\alpha$ is preferred. For the heat decay rate $\gamma$, selecting a relatively small value preserves vertex heat memory and ensures that the system remains sufficiently active. Finally, since the heat system is influenced by both external injection and internal diffusion, a moderate value of $\beta$ helps prevent external factors from becoming overly dominant.
In our implementation, the parameters are fixed to $\alpha = 0.5$, $\gamma = 0.1$, and $\beta = 0.3$.}

\revise{The convergence of the DHD model is formally established, and the
detailed proof is deferred to ~\cite{fullpaper} due to the page
limit.}

\stitle{Overlap-Centric Replica Placement.}
Once the sinking process is complete,  the pattern is assessed within the bridge subgraphs ($BS$s) of layer $L_{k}$ using a data replication gain metric. This metric guides the decision to either fully replicate the pattern within each $BS$ or to decompose it across the $BS$s in the linked cluster of the layer $L_{k-1}$ for cost minimization.


\etitle{Data replication gain.} Given that pattern $p$ sinks into $BS_y$ at layer $L_{k}$ and data item $x \in I_p$.
Suppose the cluster $\mathcal{C}$ in layer $L_{k-1}$ linked by $BS_y$ comprises the $BS$s $\{BS_1,\dots,BS_n\}$ that request $p$. We define  the gain of replicating $p$ in $BS_y$  as:
\begin{equation}\label{eq:replica_gain}
    \begin{aligned}
    &\small C^{rep}_{py} = \sum_{x \in I_p} \left(\Delta C^{(R)}_{xy} + \Delta\hat  {C}^{(A)}_{xy} - \Delta C^{(S)}_{xy} - \Delta C^{(W)}_{xy}\right),  \\
    &\text{where} \quad \small
    \Delta\hat  {C}^{(A)}_{xy} =R_{xy} \cdot \left(\lambda_1 \sum_{j = 1, BS_j \in \mathcal{C}}^n\sigma_{xyj}\right). 
    \end{aligned}
\end{equation}

Intuitively, fully replicating $p$ across the $BS$s within a cluster reduces the cost of remote reads and the association penalty, but increases storage and synchronization costs.
The gain is primarily driven by the access frequency $R_{xy}$.
For $ \Delta\hat  {C}^{(A)}$, the elimination of the second term in \equref{eq:cost_function} is due to the latency uniformity guaranteed within the cluster.
If $C^{rep}_{py} > 0$, then $p$ is fully replicated; otherwise, it is decomposed across the corresponding $BS$s.

\begin{figure}[t]
\vspace{-0.15in}
		\centering
		\subfloat[Pattern overlap]{\label{fig:pattern_overlap}
		\includegraphics[width=0.41\linewidth]{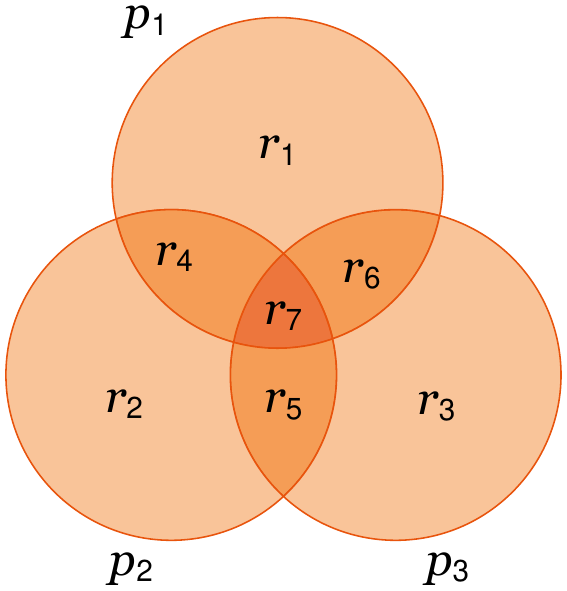}}\hspace{0.02in}
		\subfloat[Overlap-centric diffusion]{\label{fig:overlap_palcement}
		\includegraphics[width=0.5\linewidth]{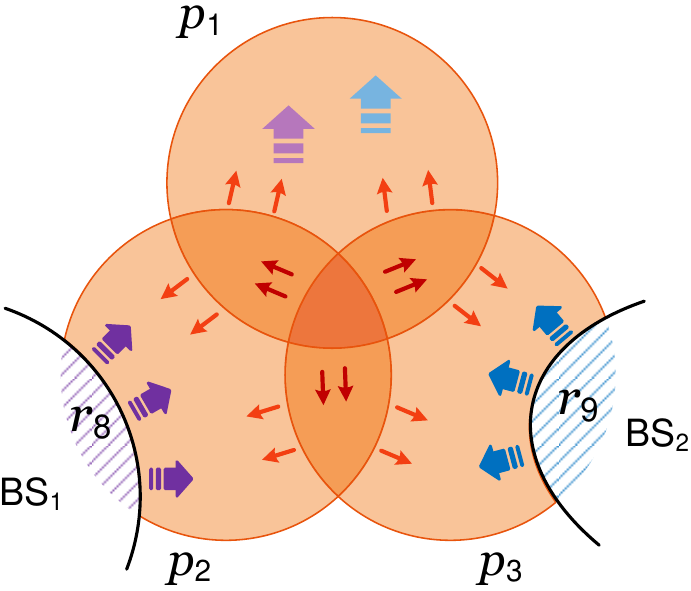}}
		\caption{Bridge subgraphs in cluster engage in overlap-centric diffusion to compete for regional placement.}
		\label{fig:overlap_centric}
		\vspace{-0.2in}
\end{figure}

\etitle{Overlap-centric regional competition.} 
We next describe the placement strategy for decomposed regions of patterns.
Due to the diversity of access patterns, overlapping regions frequently arise, where vertices and edges are shared across multiple patterns.
For example, \figref{fig:pattern_overlap} shows the maximal overlap scenario among three patterns, forming the 3-Venn diagram~\cite{ruskey2012venn}, from which all other overlap cases can be regarded as subsets.
It divides the three patterns into seven disjoint regions $\{r_1,\dots,r_7\}$, each of which is independently evaluated for placement across the $BS$s. 

Overlap regions experience a superposition effect, meaning they incur higher combined access frequencies than non-overlapping areas due to shared requests~\cite{eurosys25ohminer}. 
This amplification grows with the degree of overlap.
Accordingly, we first recompute the replication gain $\Delta C^{rep}_{r}$ for each overlapping region to decide whether it should be replicated outright or compete for placement among the $BS$s. 
To arbitrate this competition, we leverage the Directed Heat Diffusion model. 

As illustrated in \figref{fig:overlap_palcement}, the shaded regions $r_8$ and $r_{9}$ represent the overlapping parts between the patterns currently held by bridge subgraphs $BS_1$ and $BS_2$ and the three candidate patterns $p_1$, $p_2$, and $p_3$. 
We separately combine $r_8$ and $r_9$ with ${r_1, \dots, r_7}$ to form two complete heat systems for $BS_1$ and $BS_2$, respectively.
Heat updates are then performed according to \equref{eq:v_update} until the heat originating from $r_8$ and $r_9$ ceases to propagate, which occurs when there are no paths leading to regions with heat levels lower than what $r_8$ and $r_9$ can reach. 
Finally, the placement of each region is determined by comparing the magnitude of heat diffused from $BS_1$ and $BS_2$.
If the patterns co-requested by the $BS$s have no overlap with the patterns they currently hold, or if certain regions are unreachable by the heat diffused from the $BS$s, then placement is determined based on the access frequency of the $BS$s to those patterns or regions.



\stitle{Pre-Caching in High-Heat Regions.} As execution proceeds layer by layer, the final placement results are deposited in the underlying DCs. Among these results, certain regions with potentially high influence may not have been captured during the previous historical period. 

To proactively guard against sudden surges in access frequency, we use the DHD model to identify and pre-cache these high-heat regions.
First, the DC issues a global unobserved edge request to upper layers to acquire complete topological connectivity of externally cached regions.
Concurrently, it initializes a high-heat vertex set $\mathcal{O} = \{v \in V \mid \Ht_v^0 \geq \theta \}$, treating these vertices as external heat sources as defined in \equref{eq:heat_source}.
The system then iteratively updates heat states using \equref{eq:system_update} until it reaches steady state (see \equref{eq:steady_strate}).
At equilibrium, we extract the high-heat subgraph $G^H = (\Oi,\E)$, where $\Oi = \{v \in V \mid \Ht_v^* \geq \theta \}$ and $\E = \{(u,v) \mid u,v \in \Oi\}$.
This subgraph is subsequently pre-cached in the DC.
\revise{The guarantee of the lower bound on pre-caching probability is deferred to~\cite{fullpaper} due to space limitations.}

\stitle{Online Replica Eviction.} 
During online query execution, we dynamically evict replicas that become inactive over time, as caching data items that have not been accessed for a long period (\ie cold data) yields negative returns~\cite{TCC17cost}.
To identify cold segments, we model data items as a heat system: each access event creates a heat source at the corresponding item, with intensity proportional to its read frequency.
We then iteratively apply the directed heat diffusion update, allowing heat to flow through the cache topology.
Items that fail to accumulate sufficient heat are deemed cold and become candidates for eviction.
Essentially, a data item's activity level is maintained through both external heat injection and internal heat diffusion, \ie direct access incidents and indirect correlations with surrounding regions contribute to its activity.

\stitle{\revise{Update Maintenance.}}
\revise{We adopt a hybrid maintenance strategy that combines incremental updates with periodic refreshes to handle dynamic data changes.
The layered graph is a persistent, offline-built structure, with replicas cached within DCs to anticipate future accesses.
During periodic maintenance, newly inserted items that have been accessed are captured by access logs, thereby materializing request patterns. These patterns are then fed into the overlap-centric replica placement algorithm, which reacts to these pattern changes and updates replica locations accordingly.}

\revise{When a vertex or edge is deleted within a DC, maintenance starts at ${Layer}_0$.
We trigger a bottom-up cleanup: the DC handling the delete operation updates its local state, and the removal propagates upward along bridge subgraphs to affected upper-layer clusters until all replicas are removed.
In addition, the online replica eviction mechanism gradually evicts replicas as deleted data cools down, assisting the deletion process.}

%% file: 6-Routing.tex
\section{Stepwise Layered Routing}\label{sec:routing}
In this section, we present the data request routing strategies for both online and offline query modes.

\stitle{Online Query Mode.}
Built upon the overlap-centric replica placement scheme over the layered graph, \system can easily make routing decisions for online query workloads.
The core idea is to locate the required data in a layer-by-layer manner: starting from the requesting DC, the search progressively expands to increasingly distant layers until all data matching the requested pattern has been retrieved.

\begin{figure}[h!]
\vspace{-0.15in}
    \hspace{-0.1in}
    \centerline{\includegraphics[
    scale=1]{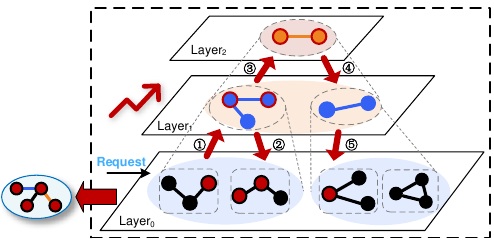}}
    \centering
    \vspace{-0.1in}
    \caption{Request routing under the online query mode. }
    \label{fig:online_routing}
    \vspace{-0.2in}
\end{figure}

For example, as illustrated in \figref{fig:online_routing}, when a pattern request is initiated by a DC in $Layer_0$,  the bridge subgraph in $Layer_1$ that links to the requesting DC is responsible for locating the missing data items within its associated cluster in $Layer_0$. The strategy greedily selects the DC that holds the largest subset of requested data items, thereby minimizing the number of DCs involved in the response.
If some data items remain unresolved,  the request is escalated to the bridge subgraph in the next higher layer, $Layer_2$, and the process continues recursively until all required data items are retrieved.

\begin{figure}[h!]
\vspace{-0.1in}
    \centerline{\includegraphics[
scale=0.89]{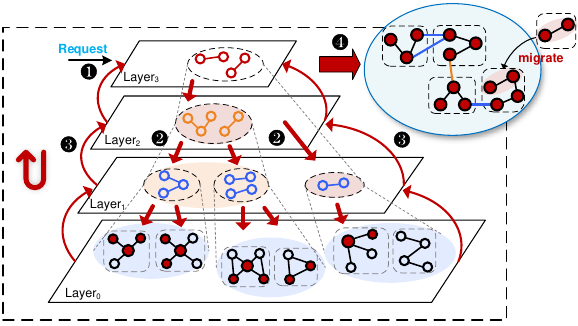}}\hspace{-0.1in}
    \centering
    \vspace{-0.3in}
    \caption{Request routing under the offline query mode. }\label{fig:offline_routing}
    \vspace{-0.1in}
\end{figure}

\stitle{Offline Query Mode.}
The offline query mode typically leverages globally distributed data to support graph analytics.
Its objective is to select a cluster of DCs that collectively cover all the required data, enabling cost-efficient execution while ensuring performance guarantees.
In contrast to the online routing strategy, the offline mode initiates data requests from the topmost layer to acquire a global view of data distribution. 
The process comprises two stages: a top-down data localization phase and a bottom-up assembly phase, as illustrated in \figref{fig:offline_routing}.
During the localization phase,  bridge subgraphs at each layer identify required data items and progressively map them to candidate DCs in a top-down manner.
Redundant cross-DC replicas are not resolved at this stage and are deferred to the assembly phase.
Once localization is complete, \system performs a bottom-up assembly process that progressively selects the most efficient data distribution from replicas at each layer, guided by communication cost.



In the bottom-up assembly phase, the process begins at $Layer_0$, where each DC evaluates its eligibility for inclusion in the analysis computing cluster.
If a DC is deemed unsuitable, the requested data it holds will be migrated. The condition for triggering data migration is defined as:
\begin{equation}\label{eq:migra_gain}
\begin{aligned}\small
    \iota \cdot s^{msg} \cdot(|I^{rep}_v| + |BS|) - \sum_{x \in I^{local}}s_x > (1-\eta_L)\cdot\xi,
\end{aligned}
\end{equation}

Here, $\iota$ denotes the predefined number of algorithm iterations, and $s^{msg}$ represents the size of messages exchanged across DCs. The terms $|I^{rep}_v|$ and $|BS|$ represent the number of replica and boundary vertices, both contributing to inter-DC communication volume. The second term captures the size of the local data that would need to be migrated if the DC is excluded.
In our implementation, $\xi$ is set to $20\%$ of the inter-DC messages volume.
$\eta_L$ denotes the ratio between the average latency of the current layer and that of the topmost layer.
To mitigate the impact of high inter-DC transmission latency on algorithm performance, the migration threshold is progressively relaxed in higher layers.
After the migration evaluation in $Layer_0$ is completed, each bridged subgraph in $Layer_1$ redistributes the migrated data by hashing it to the retained DCs within the same cluster.
If no DC is retained within a cluster, the bridge subgraph at the upper layer completes the data assembly and subsequently forwards the requested data to the designated destination that minimizes communication cost while balancing the load.
This process proceeds layer by layer in an upward manner, progressively expanding its scope to perform cross-DC data assembly, until the global result is assembled at the topmost layer.

%% file: 7-Experiment.tex
\section{Experimental evaluation}\label{sec:exper}
\subsection{Experimental Setup}\label{subsec:setup}

\begin{figure*}[tbp]
\flushleft
\vspace{-0.21in}
\hspace{-0.1in}
\begin{minipage}{.185\textwidth}\vspace{-0.03in}
\begin{center}
     \includegraphics[width=0.94\linewidth]{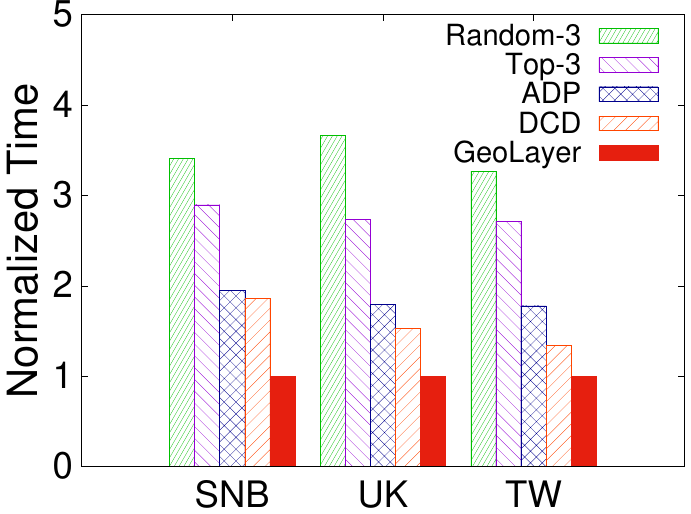}\vspace{-0.06in}
    \caption{Request latency.} \label{fig:request_latency}
\end{center}
\end{minipage}\hspace{-0.05in}
\begin{minipage}{.62\textwidth}
\begin{center}
     \includegraphics[width=0.94\linewidth]{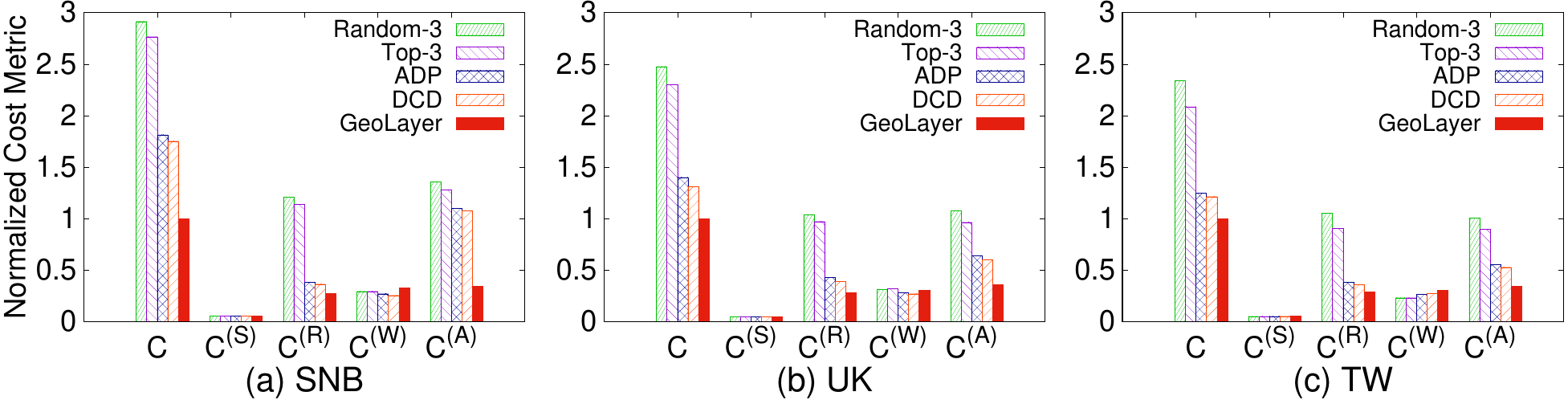}\vspace{-0.15in}
    \caption{Comparison of cost metrics.} \label{fig:cost_metric}
\end{center}
\end{minipage}\hspace{-0.05in}
\begin{minipage}{.19\textwidth}
\begin{center}
     \includegraphics[width=0.9\linewidth]{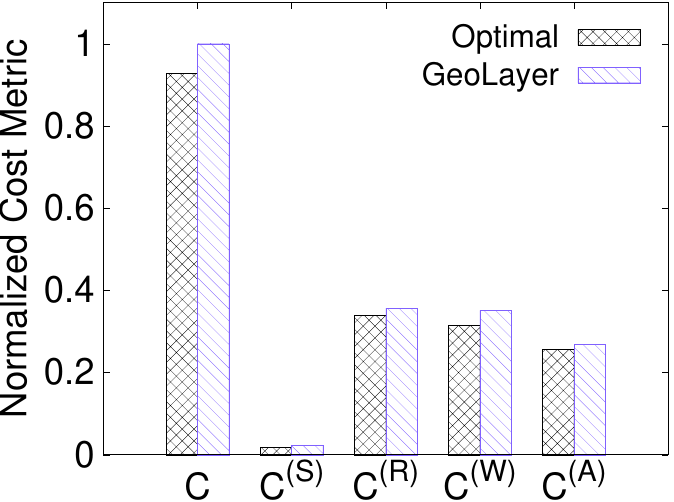}\vspace{-0.07in}
    \caption{\revise{Optimality gap.}} \label{fig:optimality}
\end{center}
\end{minipage}
\vspace{-0.2in}
\end{figure*}

\begin{figure*}[h!]
\flushleft
\hspace{-0.08in}
\begin{minipage}{.185\textwidth}
\begin{center}\vspace{0.14in}
    \includegraphics[width=0.94\linewidth]{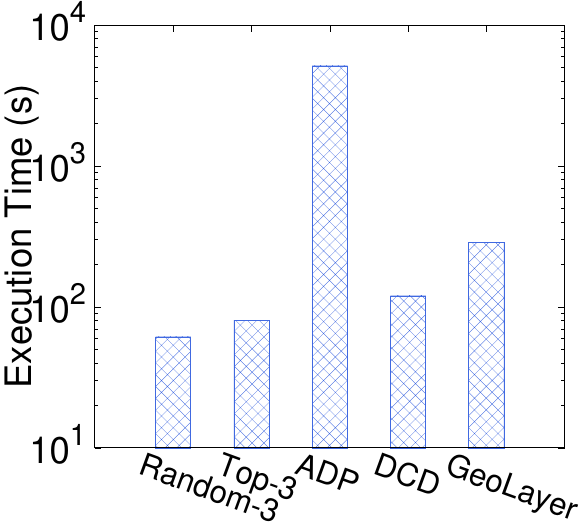}\vspace{-0.12in}
    \caption{Execution time.} \label{fig:exec_time}
\end{center}
\end{minipage}\hspace{-0.15in}
\begin{minipage}{.66\textwidth}
\begin{center}\vspace{0.08in}
     \includegraphics[width=0.94\linewidth]{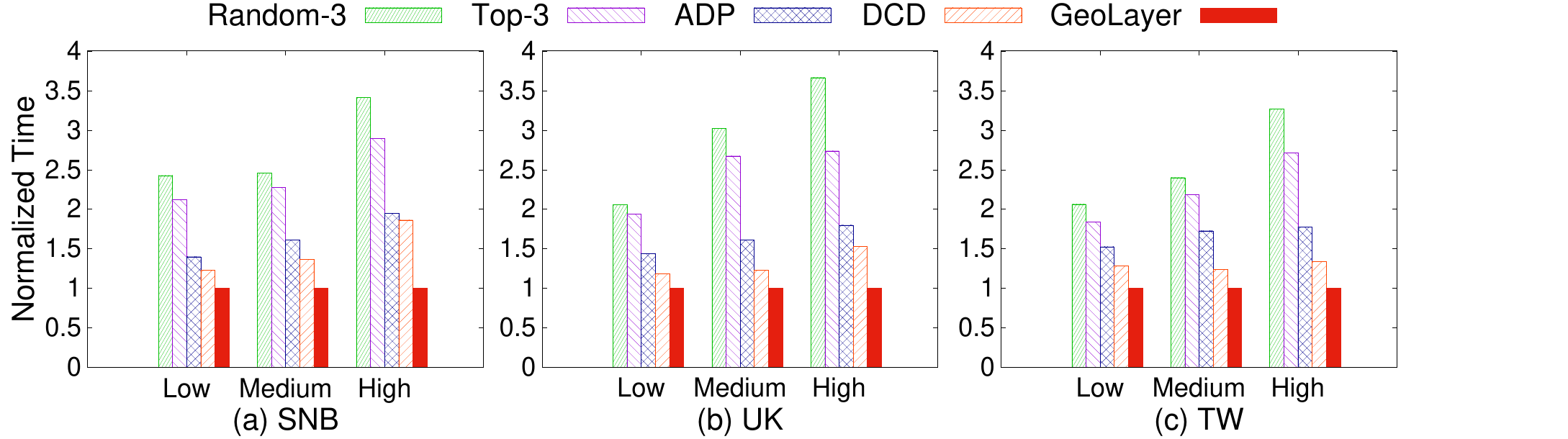}\vspace{-0.15in}
    \caption{Sensitivity to network heterogeneity.} \label{fig:hetero_network}
\end{center}
\end{minipage}\hspace{-0.65in}
\begin{minipage}{.25\textwidth}
\begin{center}
     \includegraphics[width=0.94\linewidth]{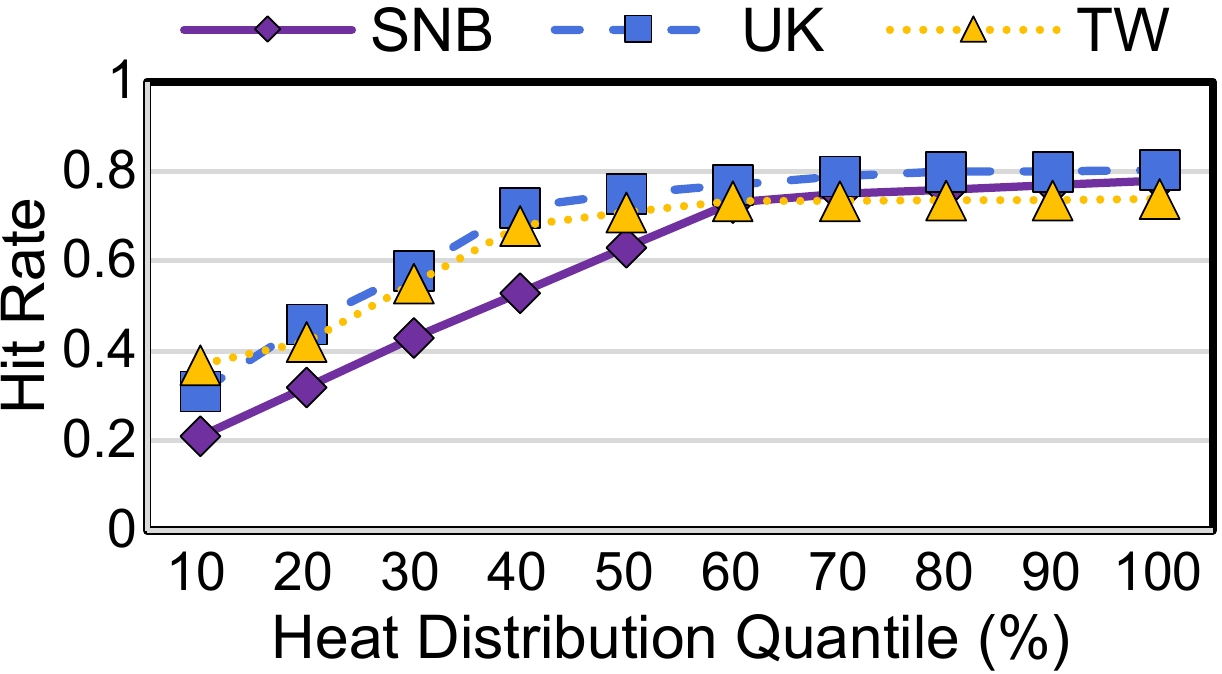}\vspace{-0.1in}
    \caption{Impact of parameter $\theta$.} \label{fig:hit_rate}
\end{center}
\end{minipage}
\vspace{-0.15in}
\end{figure*}

\stitle{Environments.}
We conduct experiments on a geo-distributed cluster deployed on Alibaba Cloud. The cluster spans eight geo-distributed data centers located in Beijing (China), Seoul (South Korea), Singapore, Sydney (Australia), Frankfurt (Germany), London (UK), Virginia (USA), and Mexico City (Mexico), with each site hosting an  \texttt{ecs.c5.xlarge}.
\revise{On this cluster, we construct the layered graph using a fixed latency interval of 100\,ms (i.e., $[0,100),[100,200),\ldots$).}
Additionally, we use the publicly advertised prices for storage, GET/PUT APIs, and data traffic on Alibaba Cloud~\cite{AlibabaOss}.

\stitle{Datasets.}
Our evaluation employs three distinct datasets: a synthetic social network generated using the LDBC-SNB benchmark with a scale factor (SF) of 10, and two real-world graph datasets, UK-2005~\cite{uk-2005} and Twitter-2010\cite{twitter-2010}. Table \ref{tab:data} provides a detailed summary of the statistics for each dataset.

\begin{table}[t]\vspace{-0.2in}
    \caption{Dataset Description}
    \vspace{-0.1in}
    \label{tab:data}
    \centering
    {\renewcommand{\arraystretch}{1.2}
    \begin{tabular}{l c c c }
        \toprule
        {\textbf{Graph}} &
        {\textbf{Vertices}} &
        {\textbf{Edges}} &
        {\textbf{Abbreviation}} \\
        \midrule
         LDBC-SNB-SF10~\cite{snb} & 36,423,181 & 231,371,359 & SNB \\
         UK-2005~\cite{uk-2005} & 39,459,925 & 936,364,282 & UK\\
         Twitter-2010~\cite{twitter-2010} & 41,652,230 & 1,468,364,884 & TW \\
        \bottomrule
    \end{tabular}
    }
    \vspace{-0.2in}
\end{table}


\stitle{Workloads.} 
We evaluate \system using two types of workloads. (i) Online graph pattern request: We use a subset of the LDBC interactive workload~\cite{sigmod15ldbc}, which defines operational queries over a social network. Specifically, we include six interactive queries \texttt{IS-1}, \texttt{IS-2}, \texttt{IS-3}, \texttt{IC-1}, \texttt{IC-2}, and \texttt{IC-3}.
In addition, we generate 3-hop random walk patterns on the \kw{UK} and \kw{TW} graphs.
We extend the graph query capabilities of RAGraph~\cite{VLDB23ragraph} by incorporating parallel Gremlin queries~\cite{nsdi21gaia}, building a geo-distributed testing platform for pattern requests.
We randomly select source vertices from different DCs and issue 1,000,000 queries to construct historical patterns with diverse latency characteristics. An additional 100,000 queries are executed as test patterns, with 30$\%$ of the data undergoing write operations.
\revise{To simulate diverse application requirements, each test pattern is assigned a random latency requirement value, which is mapped to the latency interval of one layer in the layered graph.}
(ii) Offline graph analysis: \revise{We run offline graph analytical workloads on RAGraph,  with the evaluated algorithms including \PageRank~\cite{TPDS13maiter}, Single-Source Shortest Path (\SSSP)~\cite{icde23layph}, Core Decomposition~\cite{coredocomposition},  Hyperlink-Induced Topic Search (\kw{HITS})~\cite{HITS}, and Label Propagation Algorithm (\kw{LPA})~\cite{LPA}.} The inputs are routed subgraphs from three datasets, distributed across four DCs located in Beijing (China), Singapore, London (UK), and Virginia (USA).
To reduce communication overhead, RAGraph optimizes the number of iterations. We set the number of iterations to 15 for \PageRank, 10 for \SSSP,20 for \kw{HITS}, and 10 for \kw{LPA}. 
The number of iterations for core decomposition is set to the graph’s maximum core number.


\stitle{Competitors.}
To evaluate the effectiveness of \system in the online mode, we compare it with four representative schemes:
Random-3, Top-3, ADP~\cite{TSC20ADP}, and DCD~\cite{tpds20scalable}.
Random-3 places replicas in three randomly selected DCs, while Top-3 selects the three DCs with the highest read request frequencies. Both schemes use random request routing. 
ADP~\cite{TSC20ADP} models the replica placement problem as a hypergraph partitioning task, whereas DCD~\cite{tpds20scalable} tackles it using an overlapping community detection algorithm.
For request routing, both ADP and DCD adopt greedy methods based on the set cover problem.

To assess the effectiveness of \system for offline request routing, we compare it against three baselines: RAGraph~\cite{VLDB23ragraph}, RAGraph+~\cite{TPDS24towards}, and GrapH~\cite{ICDCS16graph}.
RAGraph~\cite{VLDB23ragraph} exploits graph monotonicity to design heterogeneity-aware graph computation strategies.
We use RAGraph as the underlying execution engine and treat its original data layout as the default baseline.
In addition, it introduces a contribution-driven edge migration strategy (denoted as RAGraph+) to further optimize data distribution~\cite{TPDS24towards}. 
GrapH~\cite{ICDCS16graph} designs an adaptive edge migration strategy that considers vertex traffic and network heterogeneity.
All competitors optimize graph analysis performance by strategically selecting data locations.

\begin{figure*}[t!]
\vspace{-0.2in}
\centerline{\includegraphics[
    scale=0.32]{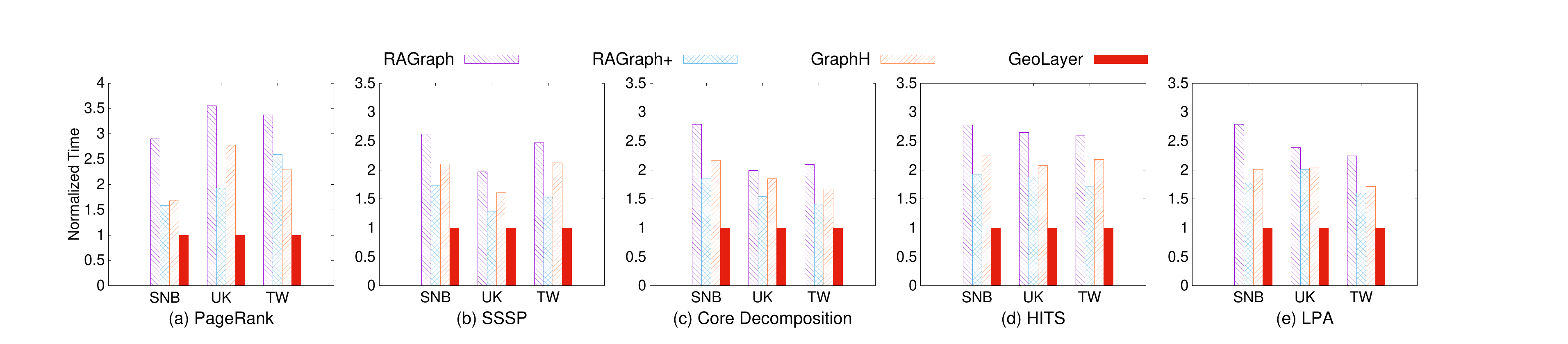}}
  \centering
  \vspace{-0.15in}
  \caption{\revise{Runtime of graph analysis.}}
  \label{fig:engine_runtime}
  \vspace{-0.15in}
\end{figure*}

\begin{figure*}[t]
\hspace{0.008in}
\centering
\begin{minipage}{.82\textwidth}
    \begin{center}
        \flushleft
        \includegraphics[width = 0.98\linewidth]{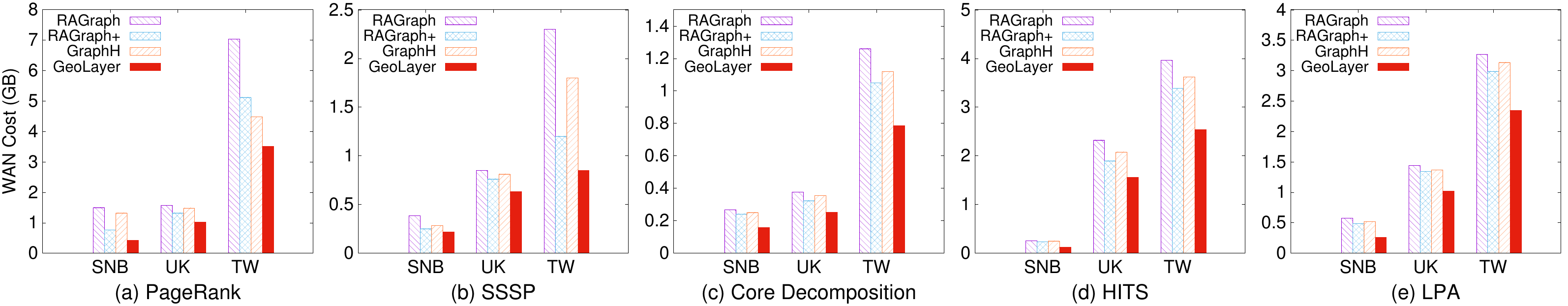}
        \vspace{-0.15in}
        \caption{\revise{WAN cost of graph analysis.}}\label{fig:engine_WAN}
    \end{center}
\end{minipage}\hspace{-0.1in}
\begin{minipage}{.18\textwidth}
    \begin{center}
         \includegraphics[width=0.94\linewidth]{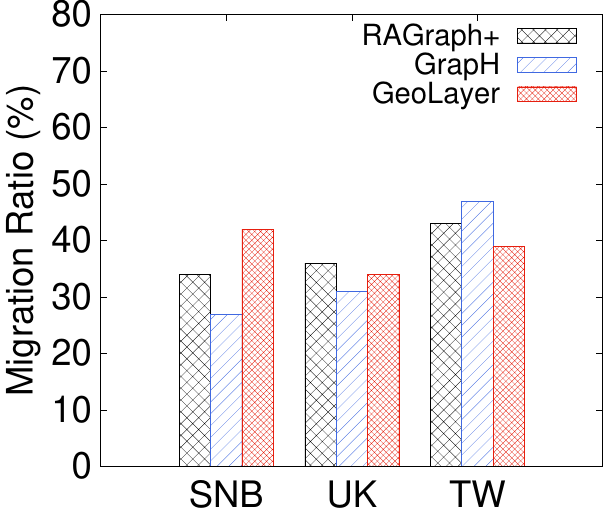}
        \vspace{-0.06in}
        \caption{Migration ratio.} \label{fig:migration_rate}
    \end{center}
\end{minipage}
\vspace{-0.2in}
\end{figure*}


\subsection{Effectiveness Analysis in Online Mode}

\stitle{Overall Performance.}
We first evaluate the request latency of test patterns under different replica placement strategies. \figref{fig:request_latency} presents the normalized response time for pattern requests across various datasets, with \system serving as the baseline (\ie normalized in unit time).
Across all datasets, \system consistently outperforms all competing approaches, achieving substantially lower response times.
Specifically, \system achieves average speedups of 3.4$\times$ over Random-3, 2.8$\times$ over Top-3, 1.8$\times$ over ADP, and 1.6$\times$ over DCD.
These improvements stem from \system's layered graph architecture, which explicitly accounts for network heterogeneity, and its replica placement strategy that is tailored to the characteristics of graph access patterns.

We also report the algorithm execution time, as shown in \figref{fig:exec_time}. Random-3 and Top-3 exhibit lower computational complexity due to their simple heuristic strategies. ADP incurs higher overhead as it relies on hypergraph partitioning and multiple rounds of iteration. Although \system involves additional complexity due to its multi-layered design, it remains efficient by leveraging cluster-level parallelism, producing results within a reasonable execution time.

\stitle{Cost Metrics.}
We further evaluate system cost metrics to understand the sources of performance improvement.
As shown in \figref{fig:cost_metric}, all objective values are normalized to the total cost ($C$) incurred by \system.
Overall, \system achieves substantially lower objective values across all datasets. On average, it reduces the cost by 60.8$\%$ (up to 65$\%$) over Random-3, 57.5$\%$ (up to 63.8$\%$) over Top-3, 31.1$\%$ (up to 44.8$\%$) over ADP, and 28.1$\%$ (up to 42.9$\%$) over DCD.
The marked reduction in association penalty cost ($C^{(A)}$) highlights \system's effectiveness in addressing network heterogeneity and minimizing inter-DC associated data access.
By incorporating the DHD model, which captures the access characteristics of graph patterns, \system significantly reduces the read cost ($C^{(R)}$).
Together, these two enhancements account for the observed reductions in request latency.
\revise{In contrast, the WAN costs of write operations ($C^{(W)}$) may occasionally rise during replica synchronization, as \system deploys additional replicas to further optimize the association penalty cost. Nevertheless, this trade-off is justified, since in practice the frequency of Write updates is typically much lower than that of Reads (\ie $C^{(R)}+C^{(A)}$), resulting in an overall net benefit.}

 
\stitle{\revise{Optimality Gap.}} \revise{To quantify the closeness of \system to the optimal solution, we employ an MILP solver (PuLP/CBC)~\cite{PuLP} on the small-scale WIKI-vote dataset~\cite{wiki-vote} (7,115 vertices and 103,689 edges). The setup involves four DCs, with 1,000 historical query patterns and 100 test patterns. We solve \equref{eq:objective_func} to obtain the global optimum ($C^{*}$) for comparison.
\figref{fig:optimality} reports the costs of \system and the global optimum across different metrics, with all values normalized to \system’s total cost ($C$).
We compute the overall optimality gap as: $\mathrm{Gap}=\frac{C-C^{*}}{C^{*}}\times 100\% = 7.8\%$.
This result shows that \system achieves small optimality gaps under the same constraints, indicating that our replication and routing schemes are near-optimal on small graphs, while scaling effectively to large deployments where exact optimization is intractable.}

\stitle{Sensitivity to Network Heterogeneity.}
We investigate the impact of network heterogeneity on system performance in geo-distributed environments.
To this end, we construct low, medium, and high heterogeneity network settings by selecting DCs from different geographical regions.
The low heterogeneity setting consists of 8 DCs located within mainland China. The medium heterogeneity setting includes 8 DCs distributed across Asia. The high heterogeneity setting involves globally distributed DCs, as described in the \ref{subsec:setup} setup.
\figref{fig:hetero_network} shows the normalized response times for graph pattern requests across all datasets.
Compared to the competitors, \system achieves an average speedup of 1.7$\times$ (up to 2.4$\times$) under the low-heterogeneity network, 1.9$\times$ (up to 3.0$\times$) under the medium-heterogeneity network, and 2.4$\times$ (up to 3.7$\times$) under the high-heterogeneity network.
These results highlight the robustness of \system's layered graph architecture in adapting to varying degrees of network diversity, particularly in environments with pronounced network variation.

\stitle{Pre-Caching Accurarcy.}
\eat{We evaluate the effectiveness of the pre-caching strategy. \figref{fig:hit_rate} reports the cache hit rates of preloaded data under varying thresholds $\theta$, corresponding to different quantiles of the vertex heat distribution.
The results show that increasing $\theta$ leads to a substantial improvement in cache hit rates across all datasets.
\system achieves near-optimal performance at around the 50$\%$-60$\%$ quantile, indicating that a small subset of frequently accessed vertices dominates the access distribution.}
To assess the effectiveness of the pre-caching strategy, we measure cache hit rates under different thresholds $\theta$, which correspond to increasing quantiles of the vertex heat distribution, as shown in \figref{fig:hit_rate}. The results indicate that selecting high-heat vertices based on a moderately high quantile (50$\%$–60$\%$) is sufficient to achieve near-optimal hit rates. This supports the observation that graph access patterns exhibit strong skewness, with a small portion of vertices being accessed disproportionately more frequently.

\subsection{Performance Comparison in Offline Mode}

\stitle{Execution Time.}
\figref{fig:engine_runtime} shows the normalized execution time of graph analysis algorithms on different datasets under each strategy.
The runtime of \system is reported as the baseline.
As can be seen from the results, \system consistently outperforms others in call cases.
\revise{Specifically, it achieves an average speedup of
2.6$\times$ (up to 3.6$\times$) over RAGraph,
1.8$\times$ (up to 2.6$\times$) over RAGraph+, and
2.0$\times$ (up to 2.8$\times$) over GrapH.}
Unlike conventional fixed-partition architectures, \system employs an adaptive routing mechanism that balances execution site selection and cross-DC communication overhead. By progressively selecting partitions in a layer-wise manner, \system effectively suppresses the formation of network heterogeneity hotspots, leading to improved efficiency.

\stitle{WAN Cost.}
We evaluate WAN usage during algorithm execution by measuring the actual volume of inter-DC data transfer.
As shown in \figref{fig:engine_WAN}, \system incurs the lowest WAN cost across all test conditions.
\revise{Specifically, \system reduces WAN cost by an average of 
42.1\% (up to 72.0\%) compared to RAGraph,
28.1\% (up to 46.8\%) compared to RAGraph+, and 
34.7\% (up to 68.2\%) compared to GrapH.}
The reduction in WAN usage stems from \system's ability to select fewer execution sites while minimizing the introduced cross–DC communication.
\system constructs $2$ execution sites for the \kw{SNB} graph, and $3$ sites for both the \kw{UK} and \kw{TW} graphs.

\stitle{Migration Ratio.}
We also report the ratio of migrated edges under each strategy, as shown in \figref{fig:migration_rate}. \system achieves a migration ratio ranging from 34\% to 42\% across the three datasets.
On the \kw{SNB} graph, \system exhibits a relatively higher migration ratio, primarily due to the consolidation of data into only two execution sites to enable more efficient processing.
All evaluated strategies consider network heterogeneity in geo-distributed environments and aim to improve graph analysis performance by reducing communication overhead through limited data movement.

\subsection{\revise{Ablation Study}}
\revise{In this subsection, we quantify the contributions of the replica placement and request routing components under both online and offline workloads across three datasets.  We adopt \kw{RP+RR} (3-replica random placement + random routing) as the baseline. To isolate component benefits, we replace one or both modules: (i) \kw{RP+SR}: random placement with stepwise layered routing (routing contribution); (ii) \kw{LP+RR}: overlap-centric replica placement with random routing (placement contribution); and (iii) \kw{LP+SR}: both overlap-centric placement and stepwise layered routing (full configuration).
For offline analytics, 
We report \PageRank as a representative algorithm.}

\begin{figure}[t]
\centerline{\includegraphics[
    scale=0.35]{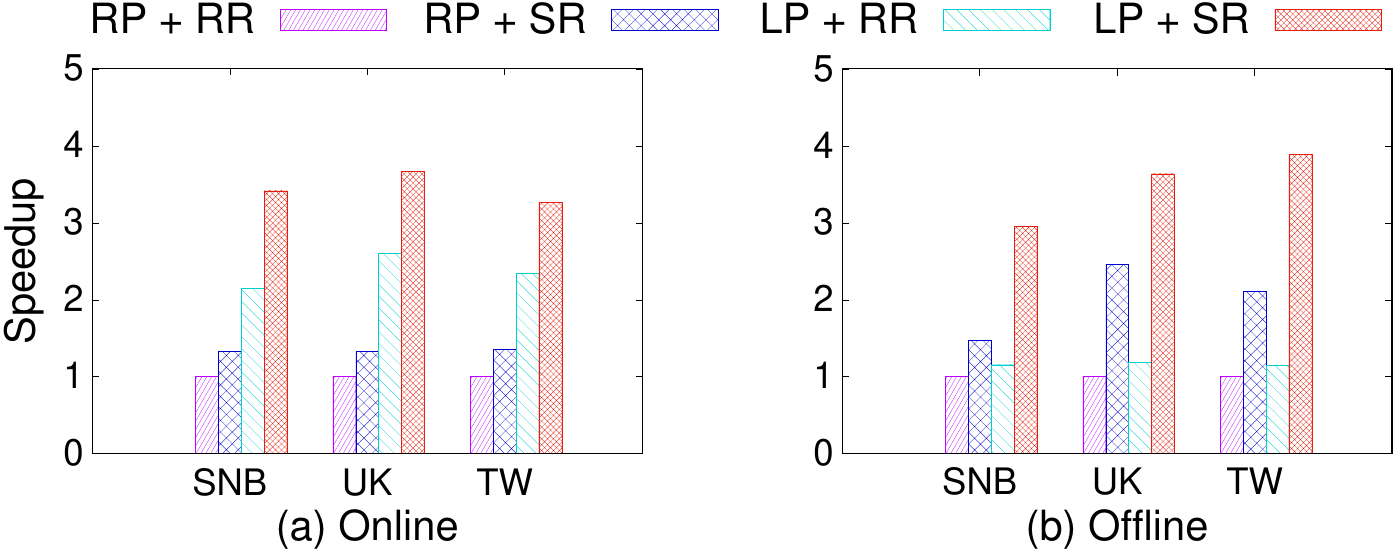}}
  \centering
  \vspace{-0.12in}
  \caption{\revise{Performance gain.}}
  \label{fig:ablation}
  \vspace{-0.23in}
\end{figure}

\revise{\figref{fig:ablation} presents speedups of all configurations relative to the \kw{RP+RR} baseline.
Replacing only the routing with \system’s stepwise layered routing (\kw{RP+SR}) yields 1.32--1.36$\times$ speedups for online workloads and 1.47--2.50$\times$ for offline workloads.
Enabling overlap-centric replica placement on the layered graph while keeping routing random (\kw{LP+RR}) achieves 2.15--2.60$\times$ speedups for online workloads and 1.15--1.19$\times$ for offline workloads.
We observe that \kw{LP} yields larger gains in the online mode, whereas \kw{SR} dominates in the offline mode. 
This is because \kw{LP} is tuned for online pattern accesses, while offline graph analytics typically involve broader, iterative access patterns that benefit more from \kw{SR}’s intra-layer locality and stepwise cross-layer escalation. Nevertheless, \kw{LP} preserves locality over random placement and thus provides consistent gains even for offline workloads.
Finally, \kw{LP+SR} (\system) improves performance to 3.26--3.66$\times$ for online workloads and 2.95--3.88$\times$ for offline workloads, indicating that both designs in \system are complementary and jointly deliver the best performance.}

%% file: 8-RelatedWork.tex
\section{Related Works}\label{sec_related}

\stitle{Data Replication and Request Routing.}
A large body of work~\cite{arxiv25skystore,ICDCS21geocol,NSDI15costlo,JSAC19writeaware,kbs21adaptive,sigmod24skypie,TCC17cost,CN19data,JGC19load,zhang2020qos,consistency19scalable,CC19consistency,luo2019Safty,SoCC18load} focuses on optimizing replica placement and request routing decisions in geo-distributed environments.
These efforts address the problem from multiple perspectives, considering factors such as replica placement~\cite{kbs21adaptive,sigmod24skypie,CN19data,TCC17cost}, caching time~\cite{arxiv25skystore}, data migration~\cite{TPDS24towards,ICDCS16graph}, QoS/QoE/SLA guarantees~\cite{ICDCS21geocol,NSDI15costlo,zhang2020qos}, consistency models~\cite{JSAC19writeaware,consistency19scalable}, security policies~\cite{CC19consistency,luo2019Safty}, and load balancing~\cite{SoCC18load,JGC19load}.
Several studies further incorporate data association into decision-making processes~\cite{infocom14multi,INFOCOM15ADP,infocom16sketch,tpds20scalable, TSC20ADP,vldb19yugong,NSDI10volley}.
Volley~\cite{NSDI10volley} proposes an iterative optimization algorithm based on weighted spherical means to handle inter-data dependencies.
Yu et al.~\cite{TSC20ADP}  model data association as a hypergraph partitioning problem to guide co-location decisions.
Liu et al.~\cite{tpds20scalable} develop a replica placement strategy based on overlapping community detection to handle associated data.
These approaches commonly abstract associated data items as edges and connect them with storage nodes to form a graph structure, thereby enabling graph-based modeling of placement decisions.
However, these associations are driven by job-level relationships, whereas \system focuses on the intrinsic structural properties of the graph.

\stitle{Geo-Distributed Graph Data Processing.} Many existing efforts~\cite{SIGIR20GeoGraph,Hotcloud18monarch,SOCC22pgpregel,TPDS19GDCN,ICDE22RLcut,TPDS25DistRLCut,tpds19GeoCut,VLDB23ragraph} aim to optimize the performance of graph-structured data processing in geo-distributed settings.
GeoGraph~\cite{SIGIR20GeoGraph} leverages clustered DCs and a metagraph to support graph query processing. 
PGPregel~\cite{SOCC22pgpregel} adopts sampling and combiner techniques to enable privacy-preserving graph processing.
Hosseinalipour et al.~\cite{TPDS19GDCN} propose an energy-efficient graph job assignment method for geo-distributed cloud networks.
RAGraph~\cite{VLDB23ragraph} exploits graph monotonicity to design heterogeneity-aware geo-distributed graph computation strategies.
These works focus on optimizing the performance of upstream computing systems, and are thus orthogonal to \system's design.

\stitle{\revise{Layered Graph Structures.}} \revise{Layered designs are widely used in graph computation and analysis. 
CompressGraph~\cite{sgimod23compressgraph} builds a two-layer traversal model via graph compression.
Layph~\cite{icde23layph} partitions a graph into an upper-layer skeleton and lower-layer disjoint subgraphs to constrain the scope of incremental computation in dynamic graph processing. 
In road networks, hierarchical structures with shortcut edges are standard for speeding up path queries~\cite{pathrouting1,pathrouting2}.
These layered approaches are tailored to the computational characteristics of graph data, but do not consider heterogeneous latency on edges.
In contrast, \system adopts a layered structure specifically designed for the latency characteristics of cross-partition edges, marking a fundamental distinction from prior work.}


\section{Conclusion}
\label{sec:conclusion}
This work contributes to optimizing replica placement and request routing for geo-distributed graph-structured data stores. The highlights of \system include: (1) We design a latency-aware layered graph structure that captures network heterogeneity and simplifies the optimization process. 
(2) We uncover distinctive pattern access features embedded in the graph topology and leverage them to develop a directed heat diffusion model that guides data placement.
(3) We propose a stepwise layered routing approach built on the layered graph, which incrementally traverses layers to locate relevant data.
Our experiment validates the effectiveness of \system.

%% file: sample.bbl

%% file: Appendix.tex
\section*{Appendix}
We first present the theoretical analysis of \system, followed by its concrete implementation.
The major notations are summarized in Table \ref{tab:notation}.

\begin{table*}[t!]
\vspace{-0.1in}
\centering
\caption{Summary of notations.}\label{tab:notation}
\vspace{-0.1in}
\begin{tabular}{c p{1.2\columnwidth}}
\toprule
{Symbol}  & {Definition} \\
\midrule
$I$, $D$, $P$, & Sets of data items ($I$), geo-distributed DCs ($D$), and read request patterns ($P$)   \\
$p$  & A read request pattern which involves multiple data items in one read transaction, $p \in P$ \\
$s_x$ & Size of data item $x$ \\

$c^{store}_d$, $c^{read}_d$, $c^{write}_d$, $c^{net}_{y d}$ 
& \textit{Costs:} storage cost at DC $d$ per unit size; read (GET) processing cost at DC $d$; write (PUT) processing cost at DC $d$; network transmission cost per unit size from DC $y$ to DC $d$ \\

$\delta_{x d}$, $\sigma_{x y d}$, $\rho_{p y d}$ 
& \textit{Binary indicators:} $\delta_{x d}$ indicates whether item $x$ has a replica in DC $d$; $\sigma_{x y d}$ indicates whether a read request for $x$ from DC $y$ is served by DC $d$; $\rho_{p y d}$ indicates whether a pattern $p$ read from DC $y$ is routed to DC $d$ \\

$R_{x y}$, $W_{x y}$, $R_{p y}$ 
& \textit{Request rates:} $R_{x y}$ is the read rate for item $x$ from DC $y$; $W_{x y}$ is the write rate for item $x$ from DC $y$; $R_{p y}$ is the read rate for pattern $p$ from DC $y$ \\

$V_d$, $E_d$, $V^B$, $E^B$
& Vertex set and edge set in DC $d$ ($V_d$, $E_d$); boundary-vertex set and cross-partition edge set ($V^B$, $E^B$) \\

$\alpha$, $\gamma$, $\beta$ & Diffusion intensity; heat decay rate;  external heat-source injection coefficient. \\

$BS$, $\mathcal{C}$ & Bridge subgraph and a cluster \\

\bottomrule
\end{tabular}
\end{table*}

\subsection{Convergence of DHD Model}
In Theorem \ref{theo:steady}, we provide a theoretical guarantee for the existence of a non-trivial steady-state solution in the heat system defined by the DHD model.

\begin{theorem}[Non-trivial steady-state existence]\label{theo:steady}
If $\alpha < \frac{\gamma}{(1-\gamma)\,\bigl\Vert L_{dir}^*\bigr\Vert}$, then the steady-state \equref{eq:steady_strate} admits a unique non-trivial solution
$\Ht^* = \beta\,(\gamma\mathbf{1} - X^*)^{-1}Q^*$, where $X^* = \alpha\,(1-\gamma)\,L_{dir}^*$, and \(\bigl\Vert L_{dir}^*\bigr\Vert\) is the induced infinity norm.
\end{theorem}

\begin{proofS}
Let \(X^* = \alpha(1-\gamma)L_{dir}^*\).  Then~\equref{eq:steady_strate} can be rewritten as
\begin{equation}\label{eq:steady_changed}
  (\gamma\mathbf{1} - X^*)\,\Ht^* \;=\; \beta\,Q^*.
\end{equation}
Since 
\(\alpha < \frac{\gamma}{(1-\gamma)\Vert L_{dir}^*\Vert}\), 
we have 
\[
\Vert X^*\Vert = \alpha(1-\gamma)\Vert L_{dir}^*\Vert < \gamma,
\]
so \(\bigl\Vert X^*/\gamma\bigr\Vert <1\).  Hence the Neumann series
\[
(\gamma\mathbf{1}-X^*)^{-1}
= \tfrac{1}{\gamma}\sum_{k=0}^{\infty}\Bigl(\tfrac{X^*}{\gamma}\Bigr)^k
\]
converges, implying \(\gamma\mathbf{1}-X^*\) is invertible.  Solving~\equref{eq:steady_changed} yields the unique solution \(\Ht^*=\beta(\gamma\mathbf{1}-X^*)^{-1}Q^*\).

Moreover, \(L_{dir}^*\) is an M-matrix, so \(\gamma\mathbf{1}-X^*\) remains an M-matrix whose inverse is entrywise non-negative.  Because \(Q^*\not\equiv0\), it follows that \(\Ht^*\not\equiv0\), i.e.\ a non-trivial steady state exists.

To see uniqueness from a fixed-point view, define
\[
\mathcal{F}(\Ht)
= \frac{X^*}{\gamma}\,\Ht + \frac{\beta}{\gamma}\,Q^*.
\]
A fixed point of \(\mathcal{F}\) satisfies \(\Ht=\mathcal{F}(\Ht)\), equivalent to~\equref{eq:steady_changed}.  For any \(\Ht_1,\Ht_2\),
\[
\small \bigl\Vert\mathcal{F}(\Ht_1)-\mathcal{F}(\Ht_2)\bigr\Vert
\;=\;\Bigl\Vert\frac{X^*}{\gamma}(\Ht_1-\Ht_2)\Bigr\Vert
\;\le\;\frac{\Vert X^*\Vert}{\gamma}\,\Vert\Ht_1-\Ht_2\Vert,
\]
and since \(\tfrac{\Vert X^*\Vert}{\gamma}<1\), \(\mathcal{F}\) is a contraction.  The Banach fixed-point theorem~\cite{bharucha1976fixed} then guarantees a unique fixed point.
\end{proofS}

\subsection{Guarantees of Pre-Caching}
We analyze the necessity of pre-caching high-heat regions.
Given  $\Oi = \{v \in V \mid \Ht_v^* \geq \theta \}$ and $\E \subseteq \Oi \times \Oi$.
Partition \(\mathcal{E}\) into observed edges \(\widetilde{\mathcal{E}}\) and unobserved edges \(\mathcal{E}\setminus\widetilde{\mathcal{E}}\).  
For each \((u,v)\in\widetilde{\mathcal{E}}\), define transition probability as
\[
\small T_{uv} = \frac{A_{uv}}{\sum_{w\in \mathcal{N}_u^{out}} A_{uw}},
\quad
T_{obs} = \min_{(u,v)\in\widetilde{\mathcal{E}}} T_{uv},
\]
so that \(T_{uv}\ge T_{obs}\).  For \((u,v)\in\mathcal{E}\setminus\widetilde{\mathcal{E}}\), impose \(T_{uv}\ge T_{min}\).

\begin{lemma}[Edge-Visit Probability Lower Bounds]\label{lm:propagation}
For any \((u,v)\in\mathcal{E}\), the visit probability 
\(\Pb_{uv} = \Ht_u^* \cdot T_{uv}\) satisfies
\[
\small\Pb_{uv} \;\ge\;
\begin{cases}
\theta \cdot T_{obs}, & (u,v)\in\widetilde{\mathcal{E}},\\
\theta \cdot T_{min}, & (u,v)\in\mathcal{E}\setminus\widetilde{\mathcal{E}}.
\end{cases}
\]
\end{lemma}
\begin{proofS}
Since \(u\in \Oi\) implies \(\Ht_u^*\ge\theta\) and \(\Pb_{uv}=\Ht_u^* \cdot T_{uv}\), substituting the respective lower bound for \(T_{uv}\) yields the result.
\end{proofS}

\begin{theorem}[Pattern-Visit Guarantee]
Let \(ph\) be a path of \(n\) edges in \(\mathcal{E}\), and define the observed-edge ratio 
\(r = \bigl|ph\cap\widetilde{\mathcal{E}}\bigr|/n.\)
Then the path visit probability satisfies 
\(
\Pb_{ph}
\;\ge\;
\jmath
\;=\;
\theta^n \cdot T_{obs}^{\,r n}\cdot T_{min}^{\,(1-r)n}.
\)
Moreover, for a pattern \(p=\{ph_1,\dots,ph_m\}\), one obtains
\(
\Pb_{p}
\;\ge\;
\prod_{j=1}^m \jmath_j.
\)
\end{theorem}
\begin{proofS}
The visit probability along a path $ph$ is
\[
\small\Pb_{ph}
= \prod_{i=1}^{n}\Pb_{u_i v_i}
= \prod_{i=1}^{n} \bigl(\Ht_{u_i}^*\,T_{u_i v_i}\bigr).
\]
Applying \lemref{lm:propagation} to each edge yields the lower bound $\jmath$.

For a pattern $p=\{ph_1,\dots,ph_m\}$, we invoke a worst‑case independence relaxation.
Since the events “visiting $ph_j$” are positively correlated, the FKG inequality implies:
\[
\small\Pb\Bigl(\bigcap_{j=1}^m ph_j\Bigr)
\;\ge\;\prod_{j=1}^m \Pb(ph_j)
\;\ge\;\prod_{j=1}^m \jmath_j.
\]
Thus $\Pb_p\ge\prod_{j=1}^m \jmath_j$, completing the proof.
\end{proofS}

\subsection{Complexity Discussion}

This section explains how each component of \system acts in concert to shrink the effective complexity of solving the objective in \equref{eq:objective_func}.

\revise{
The joint BIP formulation over the variable set 
$\{\delta, \sigma, \rho\}$ couples replica placement and routing across both DCs and query patterns, which is theoretically NP-hard. 
The goal of \system is not to compute an exact global optimum, but rather to reduce effective problem complexity through structured decomposition:}

\revise{
(i) \textbf{Latency-driven sinking shrinks candidates.} Let $\mathcal{C}_k$ be the set of clusters at layer $L_k$. The sinking operation (\algoref{alg:pattern_sinking}) confines each pattern $p$ to cluster $C\in\mathcal{C}_k$, shrinking candidates from $\lvert D\rvert$ to $\lvert C\rvert$.
Accordingly, the decision space contracts from
\[
O(|I|\,|D| + |P|\,|D|)\longrightarrow
\sum_{k}\ \sum_{C\in\mathcal{C}_k}\ O\!\big(|I_C|\,|C| + |P_C|\,|C|\big),
\]
where $I_C$ and $P_C$ denote the subsets of items and patterns confined to cluster $C$ (by sinking).
Under this clustering, global max-latency constraints are satisfied within each layer, eliminating cross-layer coupling in the optimization.
}

\revise{
(ii) \textbf{Cluster-level separability weakens objective coupling.} Within each cluster, WAN bandwidth is far more homogeneous. 
Hence, the heterogeneity-driven coupling term in \(C^{(A)}\) (Section \ref{subsec:system_model}, \equref{eq:cost_function}) is largely attenuated intra-cluster, making the total objective approximately additive across clusters and enabling parallel per-cluster solving.
}

\revise{
(iii) \textbf{Overlap-centric placement lowers granularity.} For non-replicated patterns, we further decompose them into disjoint overlap regions. 
At layer $L_k$, let $\mathcal{C}_k$ be the set of clusters. For a cluster $C\in\mathcal{C}_k$, let $BS_{C}$ denote its candidate bridge subgraphs and $\{\omega\}_\mathcal{C}$ the set of disjoint regions.
The DHD model provides a score for each region $\omega$, based on which a single-criterion greedy assignment is applied. The per-region cost is
\[
  O\big(|BS_C| + T_{\mathrm{DHD}}(C)\big),
\]
where $T_{\mathrm{DHD}}(C)$ is the time required for the diffusion to reach a steady state on cluster $C$.
Let $E_C^{B}$ be the set of cross-partition edges in $C$ used by the DHD computation and candidate filtering.
The total complexity of the placement phase is thus
\[
  O\!\left(
    \sum_{k}\ \sum_{C \in \mathcal{C}_k}
      \Big(
        |E_C^{B}| \;+\; |\{\omega\}_\mathcal{C}|\cdot\big(|BS_C| + T_{\mathrm{DHD}}(C)\big)
      \Big)
  \right).
\]
}

\revise{
(iv) \textbf{Stepwise routing prunes the search tree.}
Routing is handled in stages.
In the online mode, bottom-up greedy selection within each layer quickly prunes participating DCs by maximizing coverage before escalating to upper layers.
In the offline mode, routing first performs top-down localization and then bottom-up assembly; a local migration test (Section \ref{sec:routing}, \equref{eq:migra_gain}) replaces an exponential global subset search across all DCs, thereby avoiding combinatorial explosion.
}

\subsection{Guarantees and Decision Rules}
This section discusses how the components of \system optimize the objective function and presents the associated approximation guarantees.

\revise{Since the joint objective in \equref{eq:objective_func} is NP-hard and non-submodular, we do not claim a global constant-factor approximation to \equref{eq:objective_func}.
Instead, \system establishes: (i) feasibility by construction; (ii) monotone non-increase of a principled \emph{surrogate objective} at each actionable step; and (iii) cluster-level near-additivity, which renders subproblems separable and enables parallel optimization.}

\etitle{\revise{Feasibility.}}
\revise{Latency requirements of patterns are enforced by layering constraints. Capacity/budget checks are verified before committing replicas or routes.
Any action violating a hard constraint is rejected.}

\etitle{\revise{Monotone non-increase.}}
\revise{For each candidate action $\tau$, we evaluate the marginal change
$\Delta \widehat{C}(\tau) = \Delta C^R(\tau) - \big(\Delta C^S(\tau) + \Delta C^W(\tau) + \Delta C^A(\tau)\big)$.
Following \equref{eq:replica_gain} in replica placement and \equref{eq:migra_gain} in stepwise routing,
we accept $\tau$ only if $\Delta \widehat{C}(\tau) < 0$; otherwise $\tau$ is a no-op.
Hence every committed action is non-increasing in $\widehat{C}$, and strictly improving whenever the marginal gain is strictly negative.}

\etitle{\revise{Cluster-level near-additivity and parallelism.}}
\revise{Within a layer, bridge subgraphs induce latency-homogeneous clusters, which attenuate the heterogeneity term in $C^{(A)}$ (Section \ref{subsec:system_model}, \equref{eq:cost_function}).
Consequently, the objective is approximately additive across clusters, rendering the subproblems separable and enabling parallel optimization.}

\etitle{\revise{Approximation (surrogate-level).}}
\revise{Under standard cluster-local assumptions (monotone diminishing-returns read benefit and additive storage/write/association costs),
the regional greedy admits the classic $(1-1/e)$ approximation to the surrogate objective under a cardinality constraint~\cite{Nemhauser1978}.
We emphasize that this guarantee applies to the surrogate, not to \equref{eq:objective_func} in full generality.
}

\subsection{Implementation}
This section presents the full implementation of Overlap-Centric Replica Placement in \system. We decompose the process into three components: the pattern replica sinking operation (\algoref{alg:pattern_sinking}), the Overlap-Centric Replica Placement (\algoref{alg:overlap_replica}), and the eviction of cold replicas during online query execution (\algoref{alg:online_evict}).

\revise{\algoref{alg:pattern_sinking} describes the pattern replica sinking process.
The input pattern set $P$ is derived from historical access logs and carries request frequency and latency requirement features. The layer set $L$ is constructed using a fixed-threshold bucketing scheme (\eg 100 ms per interval).}
If layer $L_k$ meets the latency requirements of pattern $p$, then $p$ is placed into all bridge subgraphs in this layer that request $p$ (Line \ref{alg1:place}).
Otherwise, $p$ will sink to layer $L_{k-1}$ and be replicated across all bridge subgraphs that request it. At layer $L_{k-1}$, its latency requirement will then be examined to determine whether it should continue sinking. (Lines \ref{algo1:sink}-\ref{algo1:replica}).

After the pattern replica sinking is completed, \system performs the overlap-centric replica placement strategy. This strategy quantifies the benefit of data replication and leverages a directed heat diffusion model to make the final placement decisions. 
\algoref{alg:overlap_replica} outlines the replica placement process.
Initially, any external pattern $p$ retrieved at a given layer $L_k$ is included in the replication procedure if it satisfies $R_{pL_k}  > W_{pL_k}$.
The process begins with the layer-by-layer sinking of pattern replicas, guided by latency requirements (Line \ref{alg2:sink}).
Next, the data replication gain $C^{rep}_{p}$ is computed to determine whether the pattern should be fully replicated or decomposed in the next lower layer (Lines~\ref{alg2:gain_begin}--\ref{alg2:gain_end}).
If \( C^{rep}_{py} < 0 \), the pattern is decomposed across the corresponding BSs (Line \ref{alg2:gain_end}). 
For pattern decomposition, patterns are first partitioned into a set of disjoint regions based on their overlapping components (Line \ref{alg2:split}). The replication gain ${C}^{rep}_\omega$ of each region is then evaluated to determine whether the region should be replicated across $BS$s or enter a competitive allocation process (Line \ref{alg2:region_gain}). For regions subject to competition, each $BS$ executes the DHD model (Line \ref{alg2:heat}), and region ownership is determined based on the resulting heat distribution (Line \ref{alg2:compete}).
This process proceeds layer by layer until all data items reach the DCs at layer \(L_0\) (Line \ref{alg2:layerbylayer}). Finally, high-heat regions are pre-cached in the DCs at layer \(L_0\) (Line \ref{alg2:precache}).

\revise{In addition, \algoref{alg:online_evict} describes the replica eviction process during online query execution.
For each access batch, vertex access frequencies are injected into the heat system and converted into vertex heat values (Line \ref{alg3:injection}). The heat is subsequently diffused and updated via the DHD model (Line \ref{alg3:DHDModel}).
Replicas identified as cold data are evicted, and the routing information is updated accordingly (Lines~\ref{alg3:cold}--\ref{alg3:update}).}

\begin{algorithm}[t!]\small
\caption{Pattern Replica Sinking Process}
\label{alg:pattern_sinking}
\SetKwFunction{FullRep}{FullReplicate}
\SetKwFunction{place}{PlaceReplica}
\SetKwFunction{optimize}{OptimizeCost}
\SetKwFunction{Sink}{ReplicaSink}
\SetKwFunction{Record}{RecordPlacement}

\KwIn{
 Layer set $L = \{L_0, L_1, \dots, L_h\}$ with latency intervals  $L_k \mapsto[t_{k-1},t_{k})$, \revise{historical} pattern set $P$ where $\forall p \in P$ has  latency requirement $\eta_p \cdot \Gamma_{\max}$, a initial pattern distribution $\mathbb{P}^{(k)}$, $k \in \{h,\dots,1\}$
}
\KwOut{Final pattern distribution $\mathbb{P}$}

\SetKwProg{Fn}{Function}{}{}
\Fn{$\Sink()$}{
    
    \For{$k \gets h$ \KwTo $1$}{
        \While{$\mathbb{P}^{(k)} \neq \emptyset$}{
            $p \gets \mathbb{P}^{(k)}.\text{pop}()$\\
            \uIf{$\eta_p \cdot \Gamma_{\max} \in (t_{k-1},t_{k})$}{
                $\place(p, \mathbb{P}^{(k)}, L_{k})$ \label{alg1:place} \\ 
                
            }
            \Else{
             \enspace\textcolor{gray}{// \ Latency violation - sinking}\\
                $\FullRep(p,L_{k-1})$ \label{algo1:sink}\\
                $\Record(p,L_{k-1})$\\
                $\mathbb{P}^{(k-1)} \gets \mathbb{P}^{(k-1)} \cup \{p\}$\label{algo1:replica}\\
                
            }
        }
    }    
    \Return  $\mathbb{P}$ 
}
\end{algorithm}
\vspace{-0.2in}

\let\oldnl\nl
\newcommand{\nonl}{\renewcommand{\nl}{\let\nl\oldnl}}

\begin{algorithm}[ht]\small
\caption{Overlap-Centric Replica Placement}
\label{alg:overlap_replica}
\SetKwFunction{Sink}{ReplicaSink}
\SetKwFunction{Replica}{PlaceReplicas}
\SetKwFunction{Gain}{CompRepGain}
\SetKwFunction{FullRep}{FullReplicate}
\SetKwFunction{Record}{RecordPlacement}
\SetKwFunction{Split}{SplitByOverlap}
\SetKwFunction{Heat}{DHDModel}
\SetKwFunction{Partial}{PartialReplicate}
\SetKwFunction{Routing}{UpdateRoutingTable}
\SetKwFunction{Precache}{PrecachehHotRegions}
\SetKwFunction{Sample}{StratifiedSample}

\KwIn{
Layer hierarchy $L = \{L_0,...,L_h\}$, initial pattern distribution  $\mathbb{P}^{(k)} \gets \{p \in \mathbb{P}^{(k)} \mid R_{pL_k} > W_{pL_k}\}$, decomposition set $\mathbb{D}^{(k)} \gets \emptyset$, cluster set $\mathbb{C}^{(k)}$ for $L_k$, and bridge subgraph set $BS^{(k)}_p$ for requested $p$
}
\KwOut{Final data placement decisions}

\SetKwProg{Fn}{Function}{}{}
\Fn{$\Replica()$}{
    $\mathbb{P} \gets \Sink()$ \label{alg2:sink}\\ 
    \For{$k \gets h$ \KwTo $0$}{
     \nonl \underline{\textit{\textbf{Phase 1: Replication vs. Decomposition Decision.}}}\\
         \uIf{$k > 0$}{
            \ForEach{$p \in \mathbb{P}^{(k)}$}{\label{alg2:gain_begin}
                ${C}^{rep}_p \gets \Gain(p, L_k)$ \\
                \uIf{${C}^{rep}_p \geq 0$}{
                    $\FullRep(p, L_{k-1})$ \\
                    $\Record(p, L_{k-1})$\\
                    $\mathbb{P}^{(k-1)} \gets \mathbb{P}^{(k-1)} \cup \{p\}$ 
                }
                \Else{
                \textcolor{gray}{// \ For decomposition} \\
                    $\mathbb{D}^{(k-1)} \gets \mathbb{D}^{(k-1)} \cup \{p\}$ \label{alg2:gain_end}
                }
                
            }
        }
     \nonl \underline{\textit{\textbf{Phase 2: Allocation of Decomposed Regions.}}}\\
         \uIf{$\mathbb{D}^{(k)} \neq \emptyset$}{
             \ForEach{$\mathcal{C}_j \in \mathbb{C}^{(k)}$}{
                $\{r_i\} \gets \Split(\mathbb{D}^{(k)},\mathcal{C}_j)$\label{alg2:split} \\
                \ForEach{$\omega \in \{r_i\}$}{
                     ${C}^{rep}_\omega \gets \Gain(\omega, L_k)$ \label{alg2:region_gain}\\
                    \uIf{${C}^{rep}_\omega \geq 0$}{
                        $\FullRep(\omega, L_{k})$ \\
                        $\Record(\omega, L_{k})$
                    }
                    \Else{
                         \ForEach{$b \in BS^{(k)}_p$}{
                            $h^{\omega}_b \gets \Heat(b,\omega)$\label{alg2:heat}\\
                        }
                        $b^* \gets {\arg\max}_{b} h_b^\omega$ \label{alg2:compete} \\
                        $\Partial(\omega,b^*)$ \\
                        $\Record(\omega,b^*,L_{k})$\\
                    }
                    \If{$k > 0$}{
                        $\mathbb{P}^{(k-1)} \gets \mathbb{P}^{(k-1)} \cup \{\omega\}$ \label{alg2:layerbylayer}
                    }
                }
            }
         }
    }
     \nonl \underline{\textit{\textbf{Phase 3: Pre-Caching for Base Layer.}}}\\
    $\Precache(L_0, \theta)$\label{alg2:precache}
}
\end{algorithm}

\begin{algorithm}[ht]\small
\caption{\revise{Online Replica Eviction}}
\label{alg:online_evict}
\SetKwFunction{Evict}{OnlineEvict}
\SetKwFunction{Heat}{DHDModel}
\SetKwFunction{Cold}{ColdSet}
\SetKwFunction{Remove}{EvictReplica}
\SetKwFunction{Update}{UpdateRoutingTable}
\SetKwProg{Fn}{Function}{}{}

\KwIn{Cache graph $G^H = (\Oi,\E)$, vertex heat set $\Ht$, access stream $\mathcal{Q}$, and cold threshold $\theta_c$}
\KwOut{Eviction decisions}

\Fn{$\Evict()$}{
  \While{serving online queries}{
   \nonl\underline{\textit{\textbf{External Heat Injection}}}\\
    \ForEach{access $(q,\Oi_q)$ in $\mathcal{Q}$}{
      \ForEach{$v\in \Oi_q$}{
       $\Ht_v \gets  \Ht_v + R_{v}^q$\label{alg3:injection}\\
       }
    }
    $\Ht \gets \Heat(G^H,\Ht$)\label{alg3:DHDModel}\\
    \nonl\underline{\textit{\textbf{Cold Identification and Eviction}}}\\
    \ForEach{$v \in \Oi$}{ 
      \If{$\Ht_v < \theta_c$}{\label{alg3:cold}
       $\Remove(v)$\\
       $\Update(v)$\label{alg3:update}\\
        }
    }
  } 
}
\end{algorithm}